\documentclass[twocolumn,showpacs,amsmath,amstex,amssymb,mathfonts,pre]{revtex4-1}
\usepackage{graphicx}

\usepackage{tipa}
\usepackage{bm}

\usepackage{amsmath}
\usepackage{amssymb}
\usepackage{amsthm}
\usepackage{amsfonts}
\usepackage{float}
\usepackage{enumitem}
\usepackage{makecell}

\begin{document}

\title{Non-linear Dynamics, Emergent Behaviors and Controlled Expansions: Towards Effective Modeling of the Congested Traffic}

\author{Bo Yang$^1$, Xihua Xu$^{2,3}$, John Z.F. Pang$^{1}$, and Christopher Monterola$^1$}
\affiliation{$^1$ Complex Systems Group, Institute of High Performance Computing, A*STAR, Singapore, 138632.}
\affiliation{$^2$ Department of Mathematics, National University of Singapore, 119076, Singapore.}
\affiliation{$^3$ Beijing Computational Science Research Center, Beijing 100084, PR China.} 
\date{\today}

\date{\today}
\begin{abstract}
We propose a framework for constructing microscopic traffic models from microscopic acceleration patterns that can in principle be experimental measured and proper averaged. The exact model thus obtained can be used to justify the consistency of various popular models in the literature. Assuming analyticity of the exact model, we suggest that a controlled expansion around the constant velocity, uniform headway ``ground state" is the proper way of constructing various different effective models. Assuming a unique ground state for any fixed average density, we discuss the universal properties of the resulting effective model, focusing on the emergent quantities of the coupled non-linear ODEs. These include the maximum and minimum headway that give the coexistence curve in the phase diagram, as well as an emergent intrinsic scale that characterizes the strength of interaction between clusters, leading to non-trivial cluster statistics when the unstable ground state is randomly perturbed. Utilizing the universal properties of the emergent quantities, a simple algorithm for constructing an effective traffic model is also presented. The algorithm tunes the model with statistically well-defined quantities extracted from the flow-density plot, and the resulting effective model naturally captures and predicts many quantitative and qualitative empirical features of the highway traffic, especially in the presence of an on-ramp bottleneck. The simplicity of the effective model provides strong evidence that stochasticity, diversity of vehicle types and modeling of complicated individual driving behaviors are \emph{not} fundamental to many observations of the complex spatiotemporal patterns in the real traffic dynamics. We also propose the nature of the congested phase can be well characterized by the long lasting transient states of the effective model, from which the wide moving jams evolve.

\end{abstract}

\maketitle 
\section{Introduction}\label{sec_introduction}
Modeling the dynamics of the highway traffic flow has been the endeavor of researchers in many disciplines for the last fifty years\cite{dogbe, as, helbing, kernerbook}. Various different models have been proposed to describe both the free and the congested phase of the traffic flow\cite{bando,JiangR_PRE01,PengGH_PhyA13,GLW_PhyA08,kerner1, kerner2,stepan, nagel,kim,naka,xue,nishinari}. Most of these models can capture the low density free flow phase, and the wide moving jams when the density is high. Kerner\cite{kernercrit} first suggested that some essential empirical features are not captured by most of these models; there exists a ``synchronized phase" that can be distinguished by a scattering of data points covering a two-dimensional region on the flow-density plane. It is argued this phase is qualitatively different from the wide moving jams, particularly when a bottleneck at the highway is present\cite{kernerexp1,kernerexp2}. This raises the questions of the relevance of the popular general motor (GM) model classes to real traffic systems, because these models only describe a two-phase transition\cite{kernercrit}. 

Most of the three-phase models are constructed by putting in a ``synchronization gap" by hand\cite{kerner1,kerner2,kerner3}, at the cost of making the models more sophisticated with significantly more parameters. These, together with other three-phase models\cite{kimca}, reproduce the ``synchronized phase" with a multitude of steady states in the congested phase. However, Helbing et.al\cite{helbing2} pointed out that the characterization of the complex congested states of the traffic flow as a single synchronized phase is delicate, and with properly adjusted parameters some GM models \emph{can} reproduce many empirical observations at the highway bottleneck\cite{helbing2}. One should also note inhomogeneous road conditions and vehicle types\cite{helbing3}, as well as stochastic driving behaviors can  contribute to the scattering of the flow-density plot. More importantly, it is not well understood if the empirical data in the congested phase comes from equilibrium/steady traffic conditions, or from slowly evolving transient ones.

Unlike conventional physical systems, where the construction of the physical models are guided and constrained by symmetries, in traffic systems such symmetries are conspicuously lacking. Even individual components are not identical to each other: each of them responds to the interaction and environment differently, and in a time-dependent way. While constructions of the traffic models are generally guided by simplicity and the use of physically transparent parameters, there is a certain level of arbitrariness in how the model should look like due to the lack of a more fundamental guiding principle. This is one of the main reasons for a plethora of different traffic models and the controversies in the field. 

In this paper, we aim to explore such guiding principles by proposing a systematic way of constructing traffic models, in order to remove the arbitrariness in the possible forms of the models. The first part of the paper is to develop a general framework to obtain the master deterministic microscopic model, which can in principle be obtained from empirical measurements. Various simplified effective models can thus be constructed via the controlled expansions of the master microscopic model. For the next part we use the published empirical data of the A-5 North German Highway as an example to illustrate how a minimal effective model should be constructed and tuned. We first establish a list of empirical features of the traffic dynamics that are commonly reported and well-defined. This includes most of the quantitative characteristics and the qualitative spatiotemporal traffic patterns on which the ``three phase" traffic theory is based. This is followed by our proposal of a simple algorithm to construct a minimal effective model that can capture all these empirical features, with the intention that such model is the simplest possible, and it has enough predictive power for it to be useful for traffic optimization and transport engineering.

The resulting effective model is within the framework of our guiding principle and is able to capture unambiguously the empirical features that are previous thought to require the sophisticated ``three phase models"\cite{kerner1,kerner2,kerner3}, or the intelligent driver models (IDM)\cite{helbing4} that have an artificial divergence when the headway goes to zero. More importantly our effective model is surprisingly simple, providing strong evidence that the non-linear dynamics \emph{alone} is fundamental to the empirical features listed in Sec.~\ref{sec_empirical}. A careful interpretation of the emergent quantities in the system of non-linear ODEs is an important component of our algorithm, in which the parameters and functional forms in the model are no longer estimated, but well-defined by macroscopic empirical features.

The requirement that empirical features are common and well-defined does not come without a price: given the complex nature of the traffic system, there are observations of various spatiotemporal traffic patterns that are difficult to differentiate in a well-defined way. Thus the predictive power of our model, as well as our theoretical understanding, is at best ambiguous for those observations. Nevertheless, we aim to achieve a clear understanding of what can be unambiguously captured by the model and theory given the set of necessary assumptions, and to establish a useful ``reference point": when better understandings of the empirical features emerge, one can systematically extend and generalize the model.

The paper will be organized as follows: In Sec.~\ref{sec_empirical} we present a list of empirical features, defining them entirely from the experimental point of view, minimizing the tendency of pre-mature theoretical interpretations. In Sec.~\ref{sec_model} we present a general framework of systematically modeling the traffic dynamics from the most general equations of motion. All assumptions of the model are listed and connections to various microscopic models in the literature are discussed. In Sec.~\ref{sec_property} we study some of the universal properties of a class of simplified effective models, which belongs to the OV model class, and focus on the quantities emerging from the non-linear anisotropic interactions between vehicles. In particular we introduce an emergent length scale that quantifies the strength of interaction between the quasisolition structures appearing in the solutions of the OV models, leading to non-trivial cluster statistics when the constant headway solutions (or the ground states) of the OV model are randomly perturbed. This emergent scale plays an important role in tuning the model and defining the time scale in the traffic dynamics. In Sec.~\ref{sec_algorithm} we present a simple algorithm in constructing a minimal effective model based on the macroscopic empirical data, and in particular their relationship with the emergent quantities in the traffic model, and in Sec.~\ref{sec_prediction} we present the model's predictions of the quantitative empirical features, together with the characteristic spatiotemporal patterns, and compare them with the empirical observations. In Sec.~\ref{sec_conclusion} we summarize our results and discuss the outlooks of the traffic modeling.

\section{Empirical Features of Traffic Dynamics}\label{sec_empirical}

\begin{figure}

  \includegraphics[height=5cm]{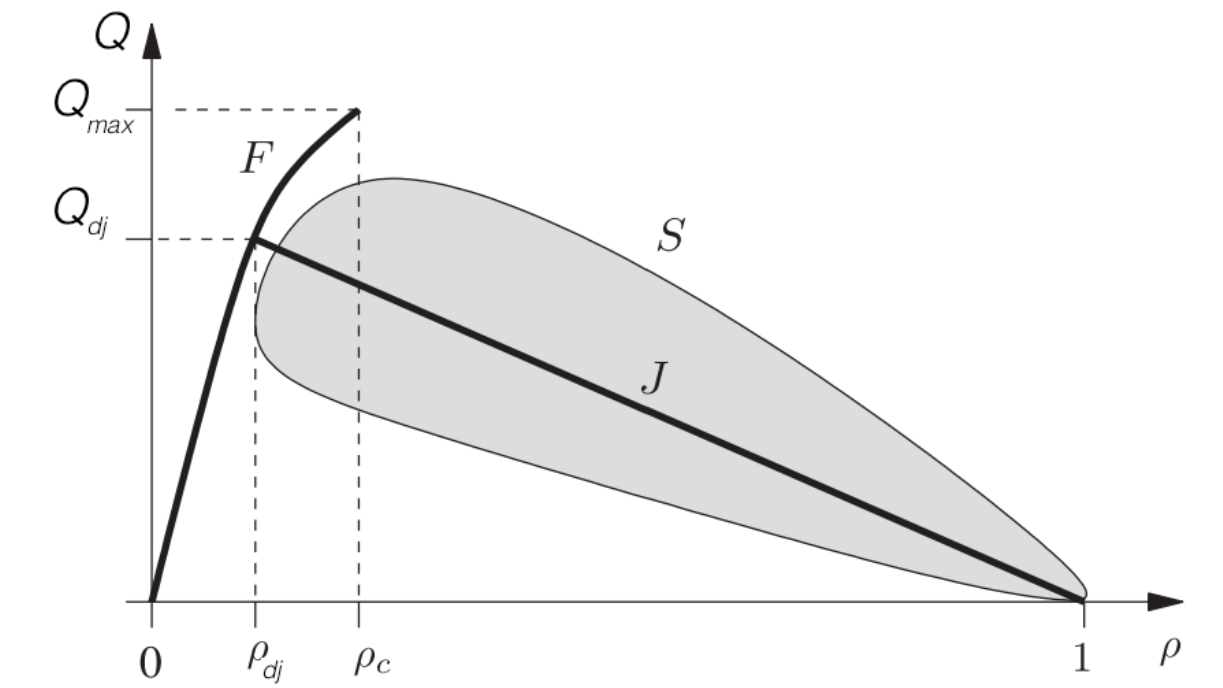}
  \caption{The schematic drawing of the flow-density plot of the highway traffic, taken from Ref.~\cite{dogbe}. The free flow phase is labeled as $F$, and the scattering of the flow-density plot in the congested phase is shaded and labeled by $S$. The Jam line labeled by $J$ describes the wide moving jam. The density is normalized by the density within a wide moving jam, and the maximum flow $Q_{\text{max}}$ is achieved at the critical density $\rho_c$. $Q_{dj}$ and $\rho_{dj}$ are the flow and density of the traffic downstream of a wide moving jam, while the gradient of the $J$ line gives the characteristic velocity of the downstream front of the wide moving jam.}
\label{schematics}
\end{figure}

Many work has been published on the empirical studies of the highway traffic dynamics, and readers can refer to \cite{kernerbook,helbing2,helbing4} for detailed information. The reconstruction of the complex spatiotemporal patterns along the highway based on the measurement of the flow and average velocity at a specific location is also a subtle issue, and an excellent discussion can be found in\cite{helbing2}. One of the most popular techniques in analyzing the traffic data is the construction of the flow-density diagram, where the flow of traffic through a fixed cross-section is plotted against the average density of the vehicles on the highway. A schematic drawing of such plot is shown in Fig.(\ref{schematics}).

It is well understood that in the limit of the road density $\rho\rightarrow 0$, the flow-density relationship is linear with very small scattering of the data. The gradient in that limit $\lim_{\rho\rightarrow 0}dQ/d\rho$ is the maximum velocity $V_{max}$ of the vehicles on the highway. This is the characteristic velocity of the vehicle when it is travelling freely with no other vehicle in the front, and is in general constrained by the speed limit, road conditions and the physical capabilities of the vehicle. 

When the density increases from zero, initially the average velocity of the vehicles does not change much, as the vehicles hardly interact with each other. Thus the flow increases linearly. At intermediate densities, the interaction between vehicles strongly reduce the average velocity, leading to a sub-linear increase of the flow against the density. This continues until the density reaches the critical density $\rho_c$, at which the flow reaches a maximum $Q_{\text{max}}$. The maximum flow $Q_{\text{max}}$ is also defined as the capacity of the highway.

In general the relationship between the flow and density becomes more scattered as the density increases towards $\rho_c$. What is more interesting is a characteristic discontinuous drop in $Q$ when $\rho$ increases past $\rho_c$, together with an onset of wide scattering of the flow-density relationship, covering a two-dimensonal area on the flow-density plane. There is clearly a phase transition here physically corresponding to the breakdown of the free flow traffic to the congested traffic. In Kerner's ``three-phase traffic theory"\cite{kernerbook}, the two-dimensional scattering of the flow-density data points corresponds to the ``synchronized flow", or the ``synchronized phase". The fundamental assumption of the ``three-phase theory" is that each scattered data point on the flow-density plane corresponds to a steady state, that can be either stable or unstable against the formation of the wide moving jam. Microscopically it is postulated that the speed adaptation of the drivers at high vehicle density leads to a non-unique relationship between the average velocity and the average headway. It is also postulated the two-dimensional covering of the ``synchronized phase" is bounded by the free flow, and a characteristic upper and lower boundary.

In this paper, we take a step back and stop short of making assumptions about the nature of the scattered flow-density relationship. The real highway traffic is an open system and it is very difficult to verify experimentally if these states are steady states, or just transient states that last for long enough time for it to be captured by the sensors. Instead we just characterize the congested phase of the traffic dynamics by a depression of the traffic flow and a much larger scattering of the flow-density relationship as compared to the free flow when $\rho<\rho_c$.

There is also a well-known ``hysteresis effect"\cite{bertini1}, whereby the traffic flow breaks down at the maximum flow $Q_{\text{max}}$, and after staying congested for an extended period of time returns back to the free flow but at a smaller flow. While due to the wildly fluctuating nature of the congested phase one cannot characterize the ``hysteresis effect" quantitatively, one can understand it from the crucial empirical observation that the traffic flow downstream of the congested phase, and in particular of the wide moving jam, is significantly smaller than the maximum flow $Q_{\text{max}}$\cite{kernerbook}, and this should also be predicted by a useful model.

Another important feature in the flow-density diagram is the ``J line", corresponding to the emergence of the wide moving jam in the traffic. The structure of the wide moving jam is surprisingly robust\cite{kernerbook}, leading to several readily measurable quantities that are characteristic of the traffic system. The gradient of the ``J line" is obtained from the characteristic velocity $V_j$ of the jam moving upstream; the intersection of the ``J line" with the free flow branch of the flow-density plot comes from the characteristic flow $Q_{dj}$,  and the density $\rho_{dj}$ of the vehicle right downstream of the wide moving jam. The intersection of the ``J line" with the density axis comes from the density of the vehicle, $\rho_j$, within the wide moving jam, where all vehicles come to a stop with zero velocity. One should also note the following relationship
\begin{eqnarray}\label{vj}
V_j=\frac{Q_{dj}}{\rho_{dj}-\rho_j}
\end{eqnarray}
It was noted by Kerner these characteristic quantities does not change with the flow and density of the vehicles away from the wide moving jam (both upstream and downstream), thus a useful traffic model should be able to capture these important quantities. 

Apart from the flow density diagram, it is also very important to characterize the rich spatiotemporal patterns of the traffic in the congested phase. It is well established that the congested phase is macroscopically rather homogeneous as compared to the wide moving jams, though microscopically one would observe spatial and temporal fluctuations of both density and velocity, sometimes in the form of numerous narrow jams in which the vehicles' velocities drop to zero momentarily\cite{kernerbook,kernerexp1,kernerexp2,helbing2,helbing4}. The distinction between the general congested traffic (the ``synchronized phase" or the ``general patterns") and the wide moving jams is rather difficult to define for the highway traffic without the presence of the bottlenecks. It is, however, unambiguous that when the traffic density exceeds $\rho_c$, one almost never see wide moving jams evolve spontaneously from the free flow, unless something drastic (for example, an accident) happens. Instead the traffic breaks down to the congested (or the ``synchronized") phase which can last up to an hour; the wide moving jams, on the other hand, emerge from the congested phase via the ``pinch effect" or merging of numerous narrow jams. This qualitative mechanism should be predicted by a useful model.

The spatiotemporal patterns of the highway traffic with the presence of bottlenecks are also very important in characterizing the traffic dynamics. The distinction between the ``synchronized flow" and the wide moving jam can be unambiguously stated by looking at the behaviors of their respective downstream front. The downstream front of the wide moving jam will move with the same characteristic velocity $V_j$ when passing through the bottleneck, while the downtream front of the ``synchronized phase" is pinned at the bottleneck\cite{kernerbook}. When the bottleneck is in the form of an on-ramp through which additional vehicles are injected into the main highway traffic, a wide moving jam passing through the bottleneck can either induce or suppress the congested flow upstream of the bottleneck, depending on the strength of the traffic flow along the main highway traffic $Q_m$ and from the on-ramp $Q_{in}$. In general the congested flow can last for a few kilometers, and the wide moving jams tend to evolve from the congested traffic and move upstream.

Detailed studies of the congested phase, including the ``general pattern" and the ``synchronized phase", also show that when the bottleneck strength increases, the mean frequency of the moving jam emergence becomes greater, and the region of the congested traffic upstream of the bottleneck is also smaller. Another important observation of the highway traffic indicates that there are still very significant fluctuations in the average velocity of the vehicles even in the region when the vehicle density is very high. These two fundamental empirical features were first pointed out to illustrate the inadequacy of the early GM model classes\cite{kernerbook}, which seem to predict exactly the opposite behaviors. They are thus very important gauges in testing the usefulness of any constructed traffic models. The empirical features listed in this section is summarized in Table.~\ref{t}.

\begin{table*}
\begin{tabular}{| c | c | c |}
\Xhline{3\arrayrulewidth}
&&Congested traffic at\\
Flow-Density Diagram & Emergence of the wide moving jams&  an on-ramp bottleneck\\
&&\\
\hline
&&\\
Pseudo-linear relationship &The congested traffic (``synchronized phase") & A wide moving jam passes \\
 when density is low&can last up to an hour & through the bottleneck \\
&&unaffected\\
&&\\
Large scattering of the congested & Wide moving jams mostly emerge & A wide moving jam may induce or \\
flow-density data points & from the congested traffic & supprese congested traffic\\
&& at the bottleneck\\
&&\\
The ``hysteresis effect" & ``Pinch effect" and the merging of &  The region of congested traffic \\
&numerous narrow jams&gets smaller with greater \\
&&bottleneck strength\\
&&\\
Significant velocity fluctuation at & & The frequency of the emergence  \\
very large vehicle density  & &of the moving jams increases with \\
&&greater bottleneck strength\\
&&\\
Quantitative features: &&\\
$V_{max}, \rho_{dj},\rho_c,\rho_{j}, Q_c, Q_{dj}$&&\\
\hline
\end{tabular}
\caption{The list of commonly observed and well-defined empirical observations from the German highway systems.}
\label{t}
\end{table*}

\section{A General framework in model construction}\label{sec_model}

A microscopic traffic model requires the understanding of the dynamics of a single vehicle based on the interaction with its environment. Unlike classical physical systems we cannot write down equations based on the symmetry of the system: there is no spatial or temporal symmetry in the traffic system, and even the interacting components are not identical to each other. In fact, the complex dynamics of the traffic system can be a result of the following (non-exhausive) factors: a). the non-linear interaction between individual vehicles; b). diversity of the vehicle/driver types; c). stochasticity of the driving behaviors; d). inhomogeneity of the traffic lanes; e). time dependence of the vehicle number and driving behaviors. To understand any empirical features observed in the real traffic, it is very important to show which one or few of those factors (and not others) are fundamentally responsible, and this should be reflected in the constructed model. From a more theoretical point of view, one would also like to understand what interesting phenomena can result from the non-linear dynamics alone, independent of all other factors.

In reality almost all highway traffics have multiple lanes, and the behavior of lane changing and overtaking can be important for certain empirical observations\cite{helbing7,herman}. In this paper, however, we ignore multiple lanes completely by modeling the traffic as a one-dimensional system. Each component, or vehicle, is labeled by a subscript $n$, which increases sequencially in the direction of the highway traffic, indicating no overtaking. We try to base our model on what we actually observe experimentally as much as possible. The acceleration of the $n^{\text{th}}$ vehicle is most generally given by
\begin{eqnarray}\label{gg}
a_n=\mathcal F_{n,\{s_i\}}\left(\{t_i\}\right)
\end{eqnarray}
where $\{s_i\}$ and $\{t_i\}$ combined  is the collection of environmental factors that influences the acceleration. The separation of these factors into groups $\{s_i\}$ and $\{t_i\}$ is arbitrary, but from the modeling perspective $\{s_i\}$ contains all the unimportant factors we would like to average over. This is because Eq.(\ref{gg}) by itself is not useful for analytic or numerical calculations. One can, however, repeatedly measure $a_n$ over a wide range of $\{s_i\}$ and $\{t_i\}$ and average over $\{s_i\}$, which is formally represented as follows
\begin{eqnarray}\label{agg}
\bar a_n=\frac{1}{N_0}\sum_{\{s_i\}}\mathcal F_{n,\{s_i\}}\left(\{t_i\}\right)=\bar f_n\left(\{t_i\}\right)
\end{eqnarray}
where $N_0$ is the proper normalization factor. If we also assume identical drivers in the traffic system, one also need to average over all the vehicles on the highway to obtain
\begin{eqnarray}\label{aagg}
\bar a_n=\frac{1}{N}\sum_{k=1}^N\bar f_k\left(\{t_i\}\right)=\bar f_0\left(\{t_i\}\right)
\end{eqnarray}
where $N$ is the total number of vehicles. One can also give a time delay on the LHS of Eq.(\ref{gg})$\sim$Eq.(\ref{aagg}) to model the reaction time of the drivers. Since the reaction time tends to be small and the acceleration is already the second time derivative, we will not consider it here. More importantly Eq.(\ref{agg}) and Eq.(\ref{aagg}) can be empirically measured after proper averaging. In practice, one does not need to know the details of $\{s_i\}$ to obtain Eq.(\ref{aagg}). A reasonably sufficient process is to record $\{a_n, \{t_i\}\}$ of many vehicles for a long period of time over diverse environments. For each set of $\{t_i\}$ one can average over the corresponding $a_n$ to obtain Eq.(\ref{aagg}).

The choice of parameters in $\{t_i\}$ is motivated physically and for convenience. Since we are interested in the spatiotemporal characteristics of the traffic dynamics, it is natural to take $\{t_i\}$ as a collections of positions and velocities. The most intuitive choice is $\{t_i\}=\{h_n,\Delta v_n,v_n\}$, where $h_n$ is the bumper to bumper headway of the $n^{\text{th}}$ vehicle, and $\Delta v_n=v_{n+1}-v_n$ is the velocity difference between the two consecutive vehicles. We are thus looking at the equation of motion of a one-dimensional system of identical components with a nearest neighbour anisotropic interaction. All other factors are averaged over and from Eq.(\ref{aagg}) we obtain:
\begin{eqnarray}\label{meqn}
a_n=f_0\left(h_n,\Delta v_n, v_n\right)
\end{eqnarray}
where for notational convenience we remove the bar representing the average taken in Eq.(\ref{agg}) and Eq.(\ref{aagg}). We will show in Sec.~\ref{sec_algorithm} that Eq.(\ref{meqn}) contains the minimal set of parameters to capture the empirical features discussed in Sec.~\ref{sec_empirical}. 

It is a data and labor intensive task to obtain empirically the exact model as defined in Eq.(\ref{meqn}), but we do know from common driving experience that $a_n$ should increase with increasing $h_n, \Delta v_n$, but decrease when $v_n$ increases. In addition, we make two key assumptions. We first assume there exist solutions to $f_0\left(h,0,v\right)=0$; the solution is basically a statement that for $h_n=h_0$, $\Delta v_n=0$, all vehicles are equally spaced apart traveling at the same velocity with no acceleration. We define this as the ground state of the traffic system at average density $h_0^{-1}$. This gives the implicit solution(s) $f_0\left(h_0,0,V^{(k)}_{op}\left(h_0\right)\right)=0$. For each $h_0$ there can exist more than one $V_{op}^{(k)}$ indexed by $k$, leading to more than one ground states with different velocities at the same traffic density.

The second assumption is that $f_0$ is smooth around $\Delta v_n=0$ and $v_n=V_{op}^{(k)}$. The two physical scales of the traffic system is $\rho_j$, the maximum vehicle density which occurs within a wide moving jam; and $V_{max}$, the maximum velocity. Using the dimensionless quantities $\tilde h_n=h_n\rho_j, \tilde v_n=v_n/V_{max}, \tilde V_{op}^{(k)}=V_{op}^{(k)}/V_{max}$ and $\Delta\tilde v_n=\Delta v_n/V_{max}$, we define
\begin{eqnarray}\label{df}
f_0=\rho_j^{-1}\kappa^2\tilde f_0\left(\tilde h_n,\Delta \tilde v_n,\tilde v_n\right)
\end{eqnarray}
where $\tilde f_0$ is also dimensionless and $\kappa=\rho_jV_{max}$. The assumption allows us to do Taylor expansion around each ground state as follows
\begin{eqnarray}\label{expansion}
\tilde a_n=&&\kappa^2\left(\frac{\partial\tilde f_0}{\partial\tilde v_n}\bigg|_{\begin{subarray}{l}\tilde v_n=\tilde V_{op}^{(k)}\left(\tilde h_n\right)\\\Delta\tilde v_n=0\end{subarray}}\left(\tilde v_n-\tilde V_{op}^{(k)}\left(\tilde h_n\right)\right)\right)\nonumber\\
&&+\kappa^2\left(\frac{\partial\tilde f_0}{\partial\Delta\tilde v_n}\bigg|_{\begin{subarray}{l}\tilde v_n=\tilde V_{op}^{(k)}\left(\tilde h_n\right)\\\Delta \tilde v_n=0\end{subarray}}\Delta \tilde v_n\right)+O(2)
\end{eqnarray}
when $\tilde v_n-\tilde V_{op}^{(k)}\left(\tilde h_n\right)$ as well as $\Delta \tilde v_n$ are small, and where $O(2)$ contains terms of higher orders of expansion. Here we also define $\tilde a_n=a_n\rho_j$, so the only dimenional scale in the equation is $\kappa$ which gives the inverse time.

A few comments are in order here. The averaging process performed in Eq.(\ref{agg}) and Eq.(\ref{aagg}) leads to a time-independent, deterministic model with identical drivers. These are the general assumptions for most microscopic traffic models in the literature. The resulting model in Eq.(\ref{meqn}) can be easily generalized to more (long-ranged) interactions, for example by including the next nearest neighbour ($v_{n+2}\in\{t_i\}$), or backward looking ($v_{n-1}\in\{t_i\}$), though we will show in Sec.~\ref{sec_algorithm} and Sec.~\ref{sec_prediction} they are not necessary.

On the other hand, it is unlikely that Eq.(\ref{meqn}) will diverge in the limit $h_n\rightarrow 0$. In fact since Eq.(\ref{meqn}) is in principle averaged from the empirical data, $a_n$ must be bounded both from the above and from below. However, popular models like the IDM with an artificial divergence when the headway goes to zero also can be expanded around its unique ground state at each average density. One would in fact expect any model with a fundamental diagram to have an equivalent model in the form of Eq.(\ref{meqn}), with a unique optimal velocity function that can capture the same physics of the traffic dynamics.

The ``speed adaptation model" in \cite{kerner3} proposes two different optimal velocity functions in two velocity ranges. If one assumes $f_0$ is analytic, this corresponds to expanding around two different $V_{op}^{(k)}$. One should note that if we assume smoothness of $f_0$ and if there exists multiple $V_{op}^{(k)}$, we would expect $\partial\tilde f_0/\partial\tilde v_n\ge 0$ when evaluated at $\Delta \tilde v_n=0,\tilde v_n=\tilde V_{op}^{(k)}$ for some $k$. Around these points Eq.(\ref{expansion}) is generally unstable with both velocities and headways diverge over time. Consequently the lowest order approximation in Eq.(\ref{expansion}) becomes invalid and most probably the system will settle into one of the stable regions where $\partial\tilde f_0/\partial\tilde v_n<0$. Thus if experimentally $f_0$ is found to be analytic with multiple $V_{op}^{(k)}$ for certain range of $h_n$, then the ``speed adaptation model" can be justified microscopically.

Most of the complicated ``three-phase" microscopic models proposed assumes that $f_0\left(h_n,\Delta v_n,v_n\right)$ is not analytic when $h_n$ is smaller than the so-called ``synchronization gap". Unlike the IDM models where the divergence of the acceleration is purely an artificial modeling tool, the non-analyticity of $f_0$ can in principle be checked with the experimental data. It would be interesting to see if the exact $f_0$ from the empirical measurement shows non-analytic behavior. In this framework the non-analyticity of $f_0$ is the necessary condition for the multitude of steady states with a non-unique relationship between the flow and density.

\begin{table*}
\begin{tabular}{| c | c |c|}
\Xhline{3\arrayrulewidth}
&&\\
Mathematical Expression & Assumptions Implemented & Comments \\
&&\\
\hline
&&\\
$a_n=\mathcal F_{n,\{s_i\}}\left(h_n,v_n,\Delta v_n\right)$ & $h_n,v_n$ and $\Delta v_n$ chosen & More (or different) parameters can be \\
&as the important parameters&chosen as important, leading to different \\
\cline{1-2}
&&empirical features captured or lost by   \\
$\bar a_n=\bar f_n\left(h_n,v_n,\Delta v_n\right)$ & The non-essential parameters in $\{s_i\}$ &averaging. While $\mathcal F$ is formal, the exact  \\
&can be averaged over&form of $\bar f_n$ can be measured experimentally\\
\hline
&&\\
$\bar a_n=f_0\left(h_n,v_n,\Delta v_n\right)$ & Identical Drivers & More than one species of \\
&&vehicles can be included\\
\hline
&&\\
$f_0\left(h_0,0,V_{op}^{(k)}\left(h_0\right)\right)=0$&At least one ground state exists& \\
&&\\
\cline{1-2}
&&Each of the assumptions\\
$\bar a_n=\sum_m\kappa_{p,q}\left(h_n\right)\left(v_n-V_{op}^{(k)}\left(h_n\right)\right)^p\Delta v_n^q$&$f_0$ is analytic around those solutions&can be experimentally verified\\
&&\\
\cline{1-2}
&Unique ground state&\\
$\bar a_n=\sum_m\kappa_{p,q}\left(h_n\right)\left(v_n-V_{op}\left(h_n\right)\right)^p\Delta v_n^q$&at each average density&\\
& (2-phase models with &\\
&a fundamental diagram)&\\
\hline
&&Higer orders can be ignored\\
$\bar a_n=\kappa_0\left(h_n\right)\left(v_n-V_{op}\left(h_n\right)\right)+g\left(h_n,\Delta v_n\right)$&Keeping the lowest order&if $v_n-V_{op}$ is small\\
& of expansion around $V_{op}$&compared to $V_{max}$ in the congested phase\\
\hline
&&\\
$\bar a_n=\kappa_0\left(v_n-V_{op}\left(h_n\right)\right)+g\left(h_n,\Delta v_n\right)$& Coefficient $\kappa_0$ independent of $h_n$&This assumption can be\\
&&experimentally verified\\
\hline
&&\\
$g\left(h_n,\Delta v_n\right)=\lambda_1\Delta v_n+\lambda_2|\Delta v_n|$&A particular form of $g$ is chosen&This is a mathematically convenient\\
&&form to include the non-linearlity of $g\left(\Delta v_n\right)$\\
\Xhline{3\arrayrulewidth}
\end{tabular}
\caption{Various stages of assumptions implemented for the construction of the effective model in Sec.~\ref{sec_algorithm}, with their corresponding mathematical expressions.}
\label{tt}
\end{table*}
\section{Emergent properties of the analytic model}\label{sec_property}

We now focus on Eq.(\ref{expansion}) with one particular $V_{op}^{(k)}$ and discuss the mathematical properties of the resulting coupled ODEs with non-linear interactions. We first rewrite Eq.(\ref{expansion}) in a simpler form, ignoring higher orders of $\left(\tilde v_n-\tilde V_{op}^{(k)}\right)$:
\begin{eqnarray}\label{meq2}
\tilde a_n=\kappa_1\left(\tilde h_n\right)\left(\tilde V_{op}^{(k)}\left(\tilde h_n\right)-\tilde v_n\right)+g\left(\tilde h_n, \Delta \tilde v_n\right)
\end{eqnarray}
here $\kappa_1\left(\tilde h_n\right)=\kappa^2\partial\tilde f_0/\partial\tilde v_n$, and the second term on the right still keeps all the higher orders of $\Delta\tilde v_n$. This is just a general form of the optimal velocity (OV) model, and its mathematical properties are quite well known. In addition to briefly discussing those properties, we will also introduce an emergent intrinsic length scale, which is not only theoretically interesting by itself, but also serves as an important tuning parameter for Sec.~\ref{sec_algorithm}.

To simplify the discussion we use the simplest form of the OV model as an example, and all the relevant properties are universal and can be qualitatively applied to Eq.(\ref{meq2}) unless otherwise stated. In this simplest case we ignore the dependence of the acceleration on $\Delta v_n$ and choose the most popular optimal velocity function\cite{bando}:
\begin{eqnarray}\label{ovf}
V_{op}^{(k)}=V_1+V_2\tanh\left(C_1\tilde h_n-C_2\right)
\end{eqnarray}
We also take the special case $\kappa_1\left(\tilde h_n\right)=\lambda\kappa^2$, independent of $\tilde h_n$. Defining $s_n=C_1\tilde h_n-C_2$, and rescaling the time $t\rightarrow\kappa C_1V_2t$, a simple transformation gives us the equivalent form
\begin{eqnarray}\label{ovmm}
\ddot s_n+\kappa_0\dot s_n=\kappa_0\left(\tanh s_{n+1}-\tanh s_n\right)
\end{eqnarray}
with the only dimensionless parameter $\kappa_0=\lambda/\left(C_1V_2\right)$. We will now focus on Eq.(\ref{ovmm}), where $s_n$ is dimensionless. The change of variable and the scaling away of the dimensions not only tells us that seemingly different driving behaviors are actually equivalent, it also makes the symmetry of ODE's in Eq.(\ref{ovmm}) explicit. One should note by definition a physical $h_n$, which is always positive, can lead to negative $s_n$ depending on the parameters in Eq.(\ref{ovf}). Linear analysis leads to a stable phase of $s_n=s_0$ against small perturbation, and the spinodal line (or the neutral stability line) is given by 
\begin{eqnarray}\label{linear}
2\text{sech}^2 s_0=\kappa_0.
\end{eqnarray}
In the regime $|s_0|>s_{c1}=|\text{sech}^{-1}\sqrt{\kappa/2}|$, a small perturbation to a uniform headway $s_0$ with $s_n(t\rightarrow 0)=s_0+\delta s_n$ leads to $s_n(t\rightarrow\infty)=s_0$, so this regime is linearly stable. The uniform headway solution is the ground state defined in Sec.~\ref{sec_model}. Note Eq.(\ref{linear}) is only exact in the limit when the perturbation goes to zero; close to the spinodal line, the uniform headway configuration is metastable, a large enough perturbation will also lead to the formation of clusters.

\begin{figure}
  \centering
  \setbox1=\hbox{\includegraphics[height=7cm]{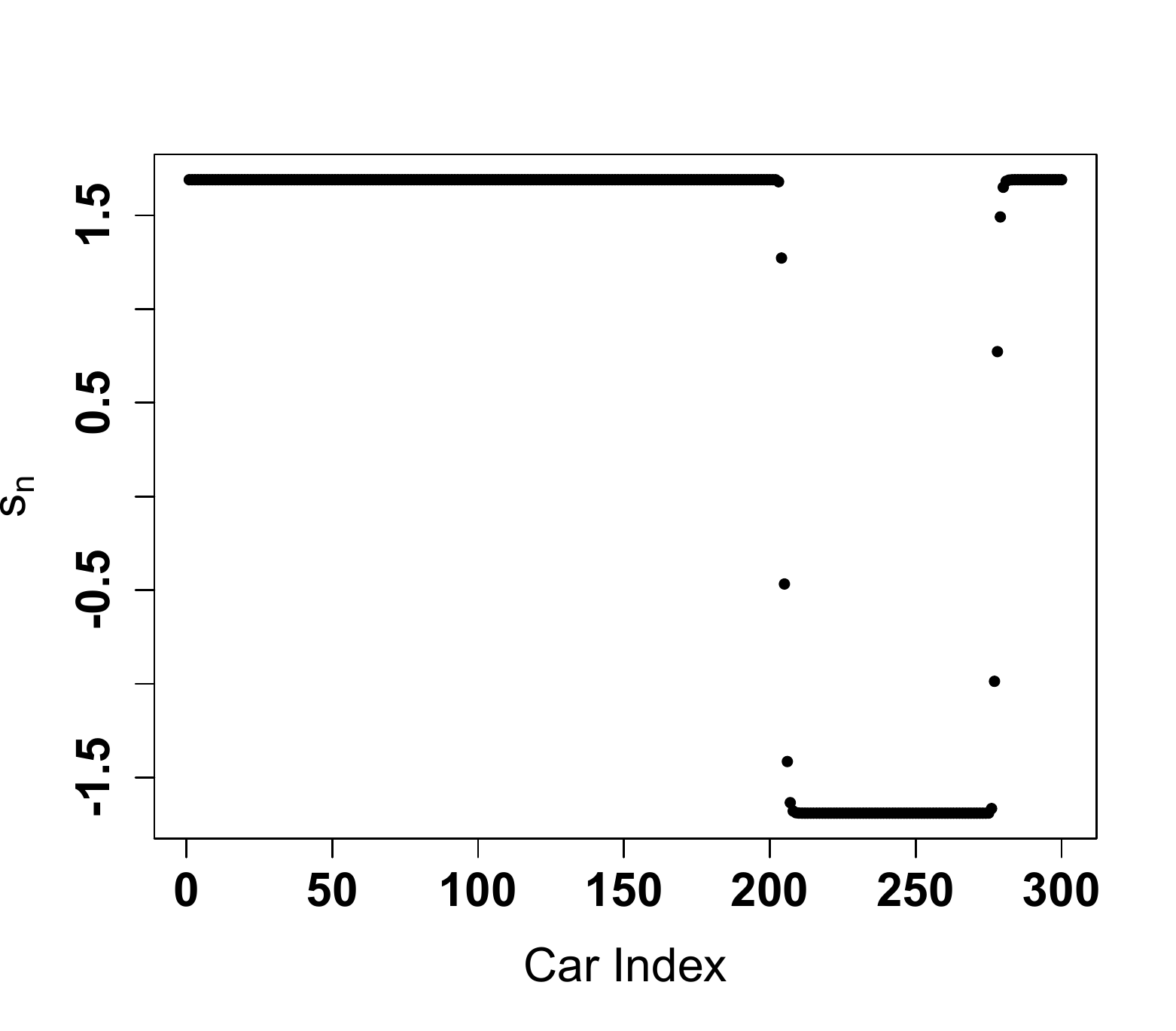}}
  \includegraphics[height=8cm]{single_peak.pdf}{\llap{\makebox[\wd1][l]{\raisebox{1.45cm}{\includegraphics[height=3cm]{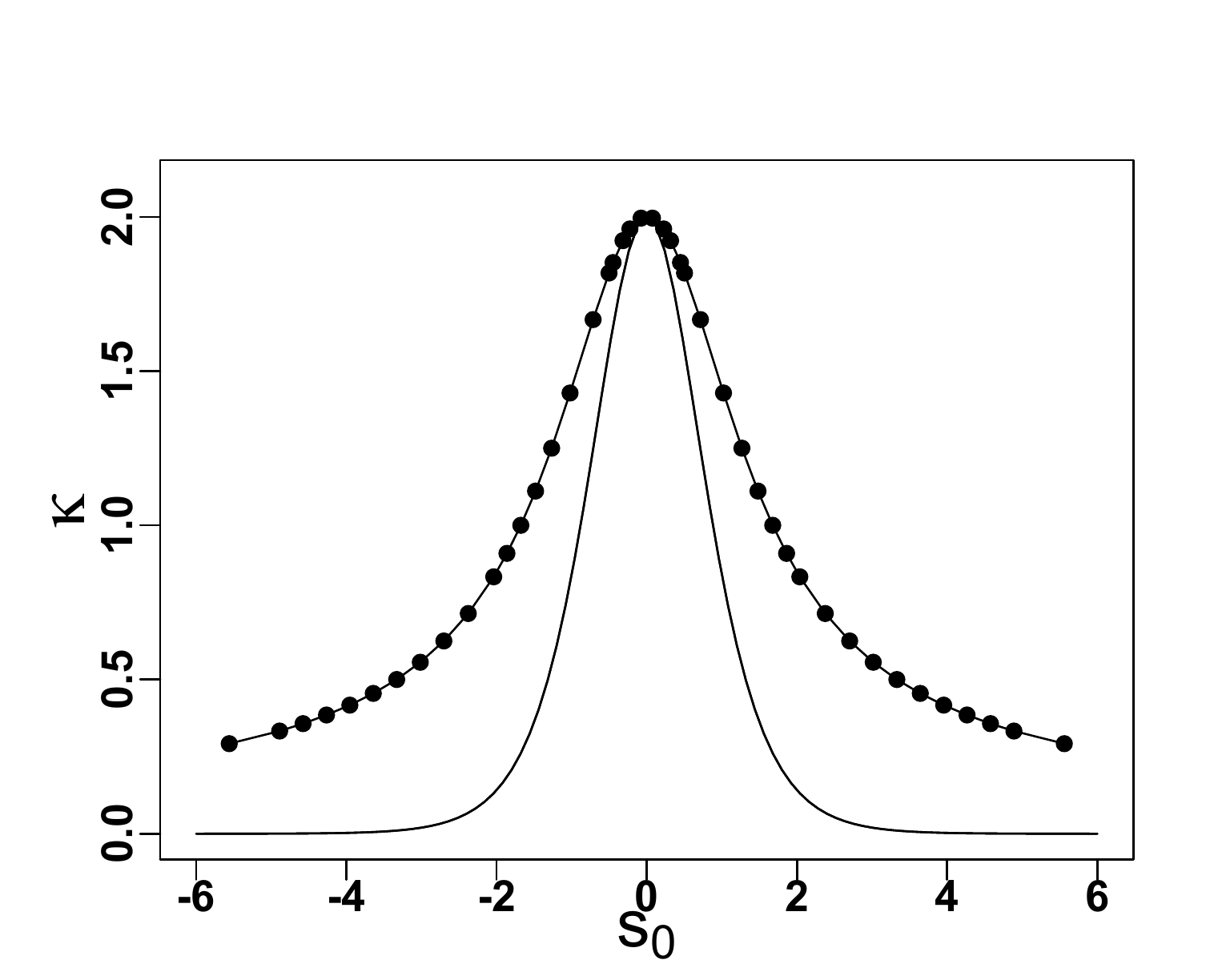}}}}}{\llap{\makebox[\wd1][l]{\raisebox{3.95cm}{\includegraphics[height=3cm]{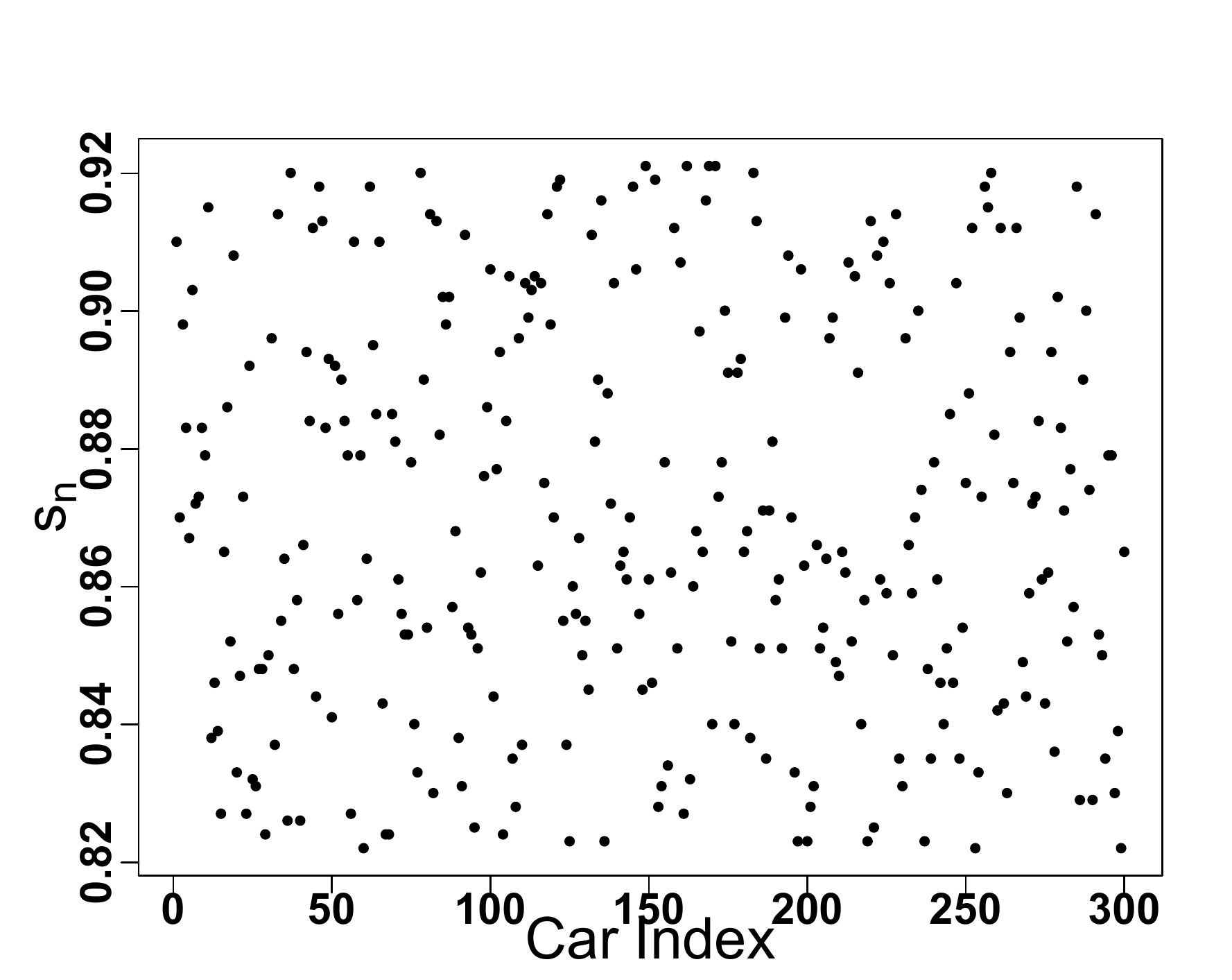}}}}}
  \caption{The plot of the headway as the function of the vehicle index, when a jam or a cluster is formed. This cluster configuration evolves from a random initial headway distribution, as shown in the top inset. The bottom inset is the spinodal curve (the solid line without circles, plotted from Eq.(\ref{linear})), and the coexistence curve from the numerical calculations (the solid line fitting the solid circles). The solid circles are numerically observed extremum headways at different $\kappa_0$.}
\label{singlepeak}
\end{figure}

We now show that the coexistence curve that separates the metastable phase and the absolutely stable phase can be numerically obtained from the cluster structure. Firstly, in the regime $|s_0|<s_{c1}$, it is well known that small perturbations will grow in time with the formation of clusters, as shown in Fig.(\ref{singlepeak}), where a random initial condition settles into a configuration with the majority number of vehicles having two extremum headways given by $\pm s_{c2}$. As smaller $s_n$ implies higher physical vehicle density, vehicles with headway $-s_{c2}$ form clusters or jams of very high density with minimal velocity, while vehicles with headway $s_{c2}$ moves with very high velocity, forming anti-clusters. Interestingly like $s_{c1}$, the numerical value of $s_{c2}$ only depends on $\kappa_0$ but \emph{not} on $s_0$, so the cluster structure is unique once the parameters in the model is fixed.

Secondly the number of vehicles involved in the ``kink" or ``anti-kinks" are independent of $s_0$ and the total number of vehicles $N$. A ``kink" is the ``go front", or the transition region from a cluster with $s_n\sim-s_{c2}$ to an anti-cluster with $s_n\sim s_{c2}$, while an "anti-kink" is the ``stop front", or the transition region from an anti-cluster to a cluster. Thus for large $N$  we can ignore vehicles in the ``(anti-)kink", and the number of vehicles in the cluster is given by
\begin{eqnarray}\label{jnumber}
N_j=\frac{N}{2}\frac{s_{c2}-s_0}{s_{c2}}
\end{eqnarray}
Clearly for $s_0\ge s_{c2}$, no clusters can be formed, given random initial perturbations of any magnitude, as long as $\sum_n\delta s_n=0$. Similarly, no anti-clusters can exist for $s_0<-s_{c2}$. We thus identify $s_{c2}$ as the coexistence curve \cite{kerner,yu,naka} and plot it together with $s_{c1}$ in Fig.(\ref{singlepeak}). The numerically calculated coexistence curve and the spinodal line coincides at the critical neutral stability point located at $s_0=0, \kappa=2$, agreeing with the previous analysis\cite{nagatani,xue}. Note that $s_n$ can be negative, and the physical vehicle density is calculated from the model parameters $C_1$ and $C_2$. There is also a duality between $s_0\leftrightarrow -s_0$, where clusters at $s_0$ corresponds to anti-clusters at $-s_0$, and all behaviors at $s_0$ are identical to those at $-s_0$. This symmetry is entirely due to the fact that the RHS of Eq.(\ref{ovmm}) is odd. In the more general model of Eq.(\ref{meq2}), this duality can be broken and it is thus not universal.

Progresses have been made in treating non-linear ODE describing car-following models analytically\cite{nakashini, yokokawa1,naka,nagatani}; For Eq.(\ref{ovmm}) it is generally accepted that one can do a controlled expansion near the critical neutral stability point and close to the neutral stability line; the former leads to the modified KdV equations plus correction terms, that gives the approximate ``(anti-)kink" solutions; the latter reduces the original model to the KdV equations plus corrections that give rise to soliton solutions\cite{kurtze}. However, away from the neutral stability line, it is clear from the numerical calculation that if one makes the vehicle index continuous, the transition between the two extremum headways is discontinuous and analytically intractable. 

One can, however, show that the ``kink" and ``anti-kink" of a single cluster move at the same velocity, by taking $s=\sum_{n=i}^js_n$. For the ``kink", the $i^{\text{th}}$ vehicle is located in the cluster, while the $j^{\text{th}}$ vehicle is located in the anti-cluster. From Eq.(\ref{ovmm}) we have
\begin{eqnarray}\label{calcv}
\ddot s+\kappa_0\dot s=2\kappa_0\tanh s_{c2}
\end{eqnarray}
The relevant set of solutions is $s=2\tanh (s_{c2})t+C$, where $C$ is an unimportant constant of integration. This gives the velocity of the ``kink" as \emph{the number of vehicles per unit time} as follows
\begin{eqnarray}\label{kinkv}
v_{k}=\frac{\tanh s_{c2}}{s_{c2}}
\end{eqnarray}
The velocity of the ``anti-kink" is calculated similarly, thus $v_k$ gives the velocity of the cluster, which again is \emph{independent} of the vehicle density of the traffic lane. Here we make the assumption that for vehicles far away from the ``kink" or the ``anti-kink", their headway takes the value of $\pm s_{c2}$. More importantly, if we concatenate two clusters together, as long as the assumption holds (e.g. when the two clusters are far away), they will move at the same velocity and will never merge.

\begin{figure}
  \centering
  \setbox1=\hbox{\includegraphics[height=3.9cm]{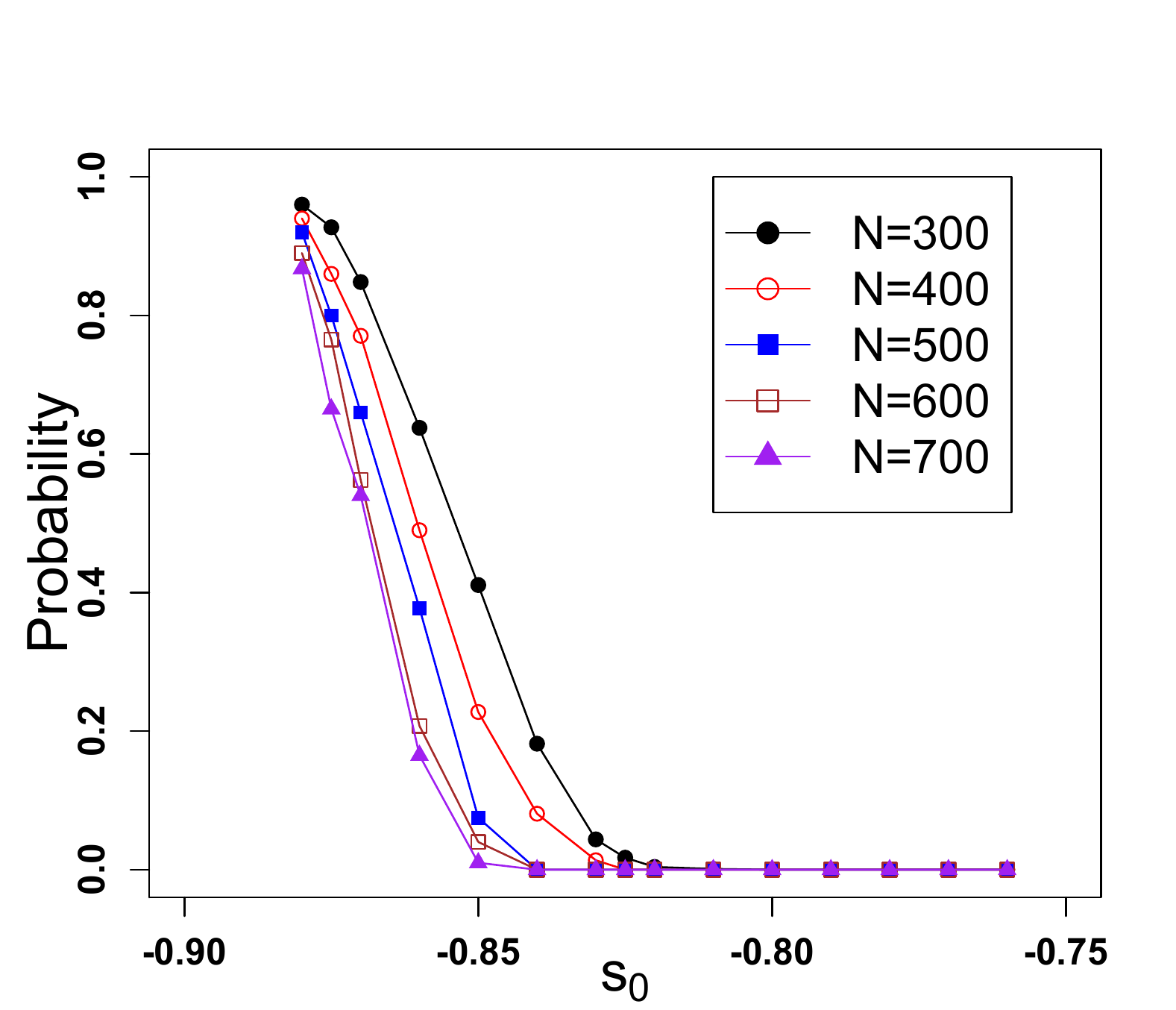}}
  \includegraphics[height=7.5cm]{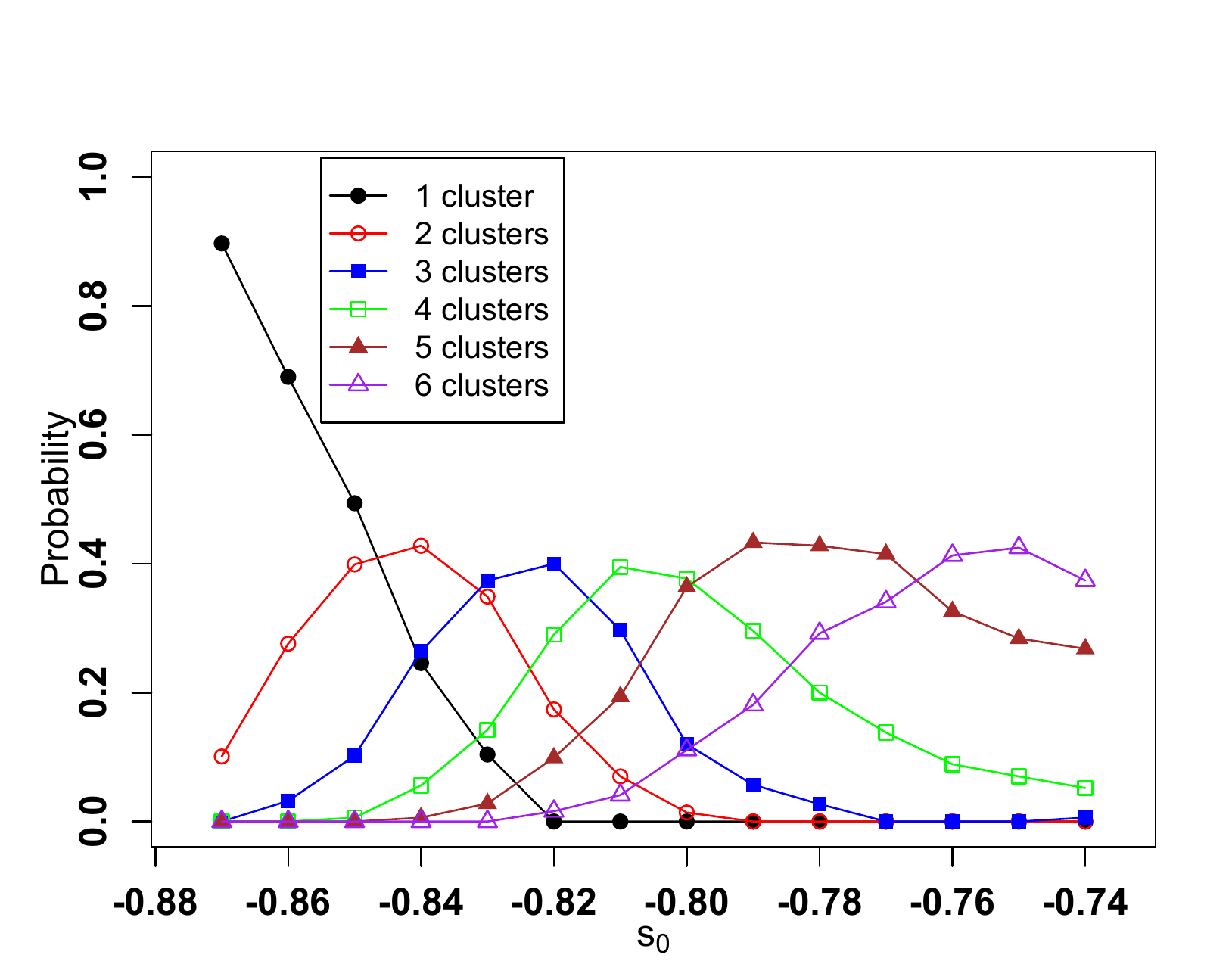}{\llap{\makebox[\wd1][l]{\raisebox{3.2cm}{\includegraphics[height=3.6cm]{probability2.pdf}}}}}
  \caption{(Color online) The probability of having one to six clusters in a single traffic lane, plotted as the function of the initial average headway $s_0$. The probability is calculated with three hundred vehicles and random initial headway perturbation. Inset: The probability of having only one cluster, as the function of the initial average headway. The probability is calculated for three hundred vehicles to seven hundred vehicles, showing that in the limit of large number of vehicles, the probability curve converges to a well-defined limit. The probability is calculated at $t=30000 s$.}
\label{probability}
\end{figure}

One interesting universal aspect of the OV models is the non-trivial probability distribution of the number of cluster formations, when the ground state is randomly perturbed. One might expect that a random initial state like the inset of Fig.(\ref{singlepeak}) should lead to a random number of clusters\cite{zhang}, at least in the limit of large $N$, subjecting to the constraint of Eq.(\ref{jnumber}). However, our numerical results show that the probability distribution of the number of clusters is not random; it strongly depends on the initial headway $s_0$ and $\kappa_0$. An example of the probability distribution is calculated and presented in Fig.(\ref{probability}), by fixing the strength of the initial random perturbation and $\kappa_0$ in Eq.(\ref{ovmm}), and only vary the initial headway $s_0$. For each value of $s_0$, sufficiently large number of random initial states are generated until the probability for each number of clusters converges. 

A few comments are in order here before we give a proper explanation. In Fig.(\ref{probability}) we only plot the part where $s_0$ is negative, because the probability distribution is \emph{identical} for $s_0$ and $-s_0$. For $|s_0|>0.87$ we can see the final state is dominated by the one-cluster configuration, and this is true even for an infinitely long traffic lane as $N\rightarrow\infty$; in this case, most probably one very large cluster develops, instead of several clusters with smaller lengths. As $|s_0|$ decreases, the probability of having more than one cluster increases, and for $|s_0|<0.82$, it is almost impossible to have just one cluster. As $|s_0|$ further decreases towards \emph{zero}, the average number of clusters most probably will tend to infinity. This cannot be observed numerically for a finite number $N$, since at $s_0=0$ the total number of vehicles in the clusters is $\sim N/2$(see Eq.(\ref{jnumber})). 
 \begin{figure}
  \centering
  \setbox1=\hbox{\includegraphics[height=6.3cm]{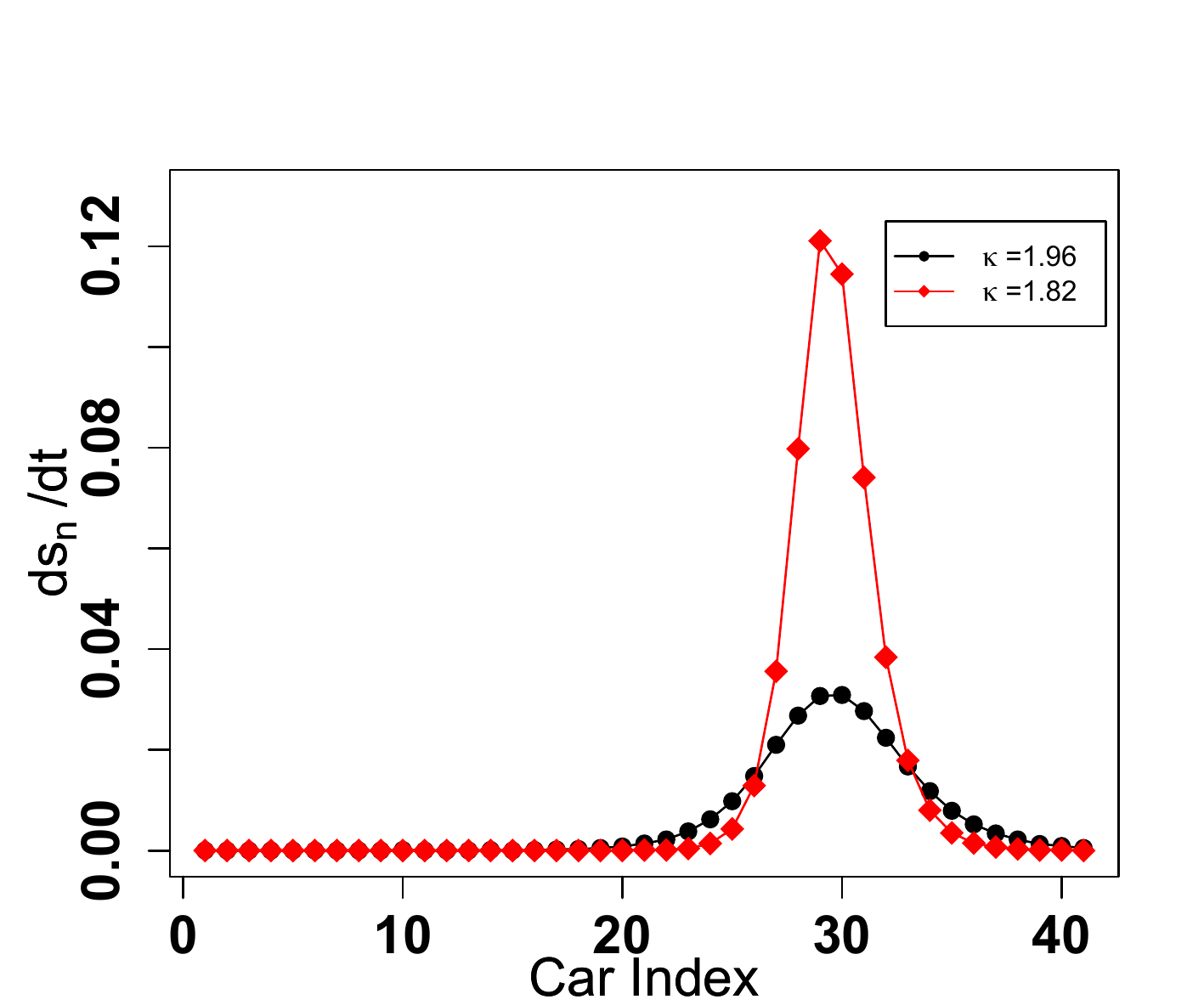}}
\includegraphics[width=9cm]{bump.pdf}{\llap{\makebox[\wd1][l]{\raisebox{1.2cm}{\includegraphics[height=3cm]{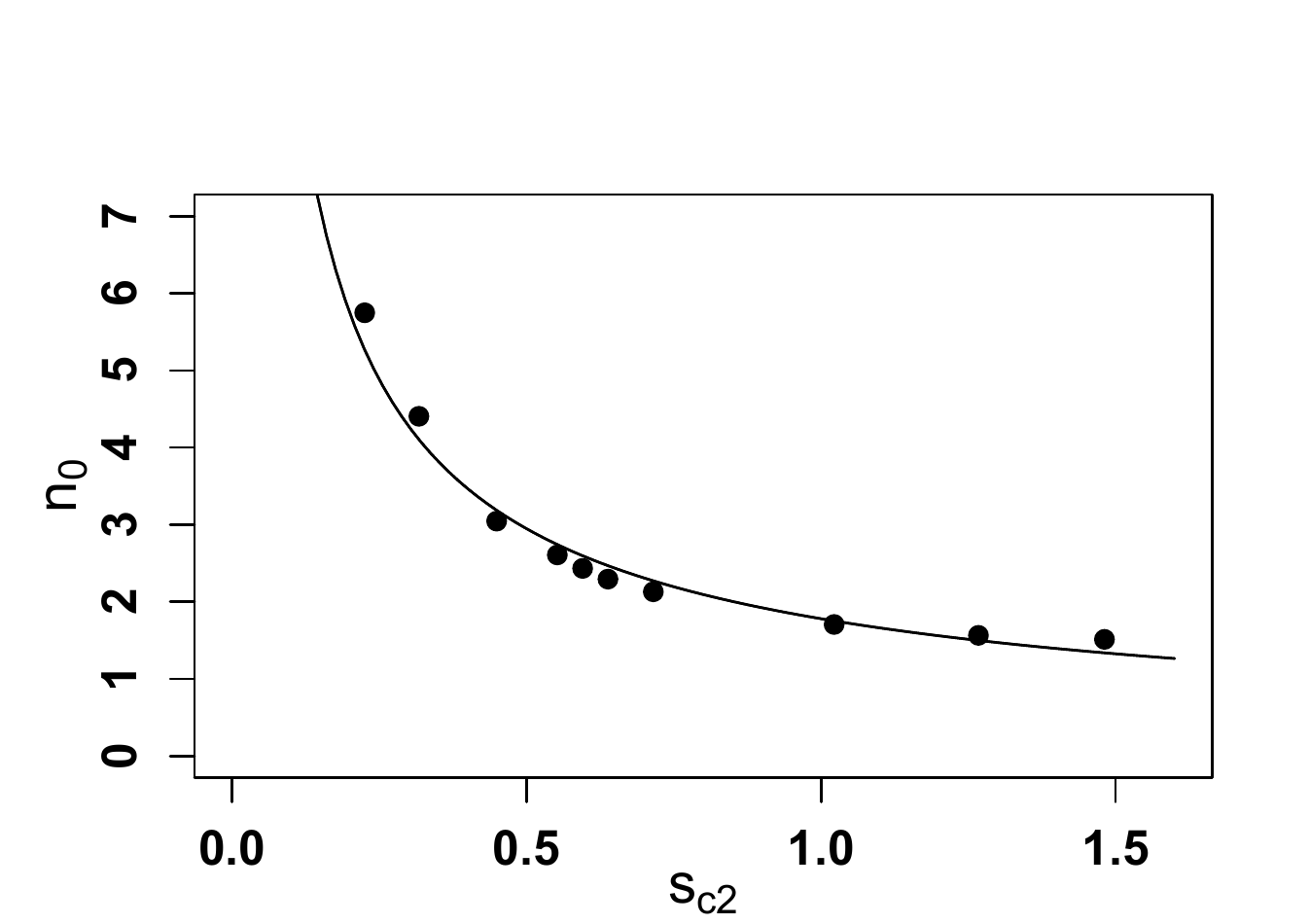}}}}}{\llap{\makebox[\wd1][l]{\raisebox{3.7cm}{\includegraphics[height=3cm]{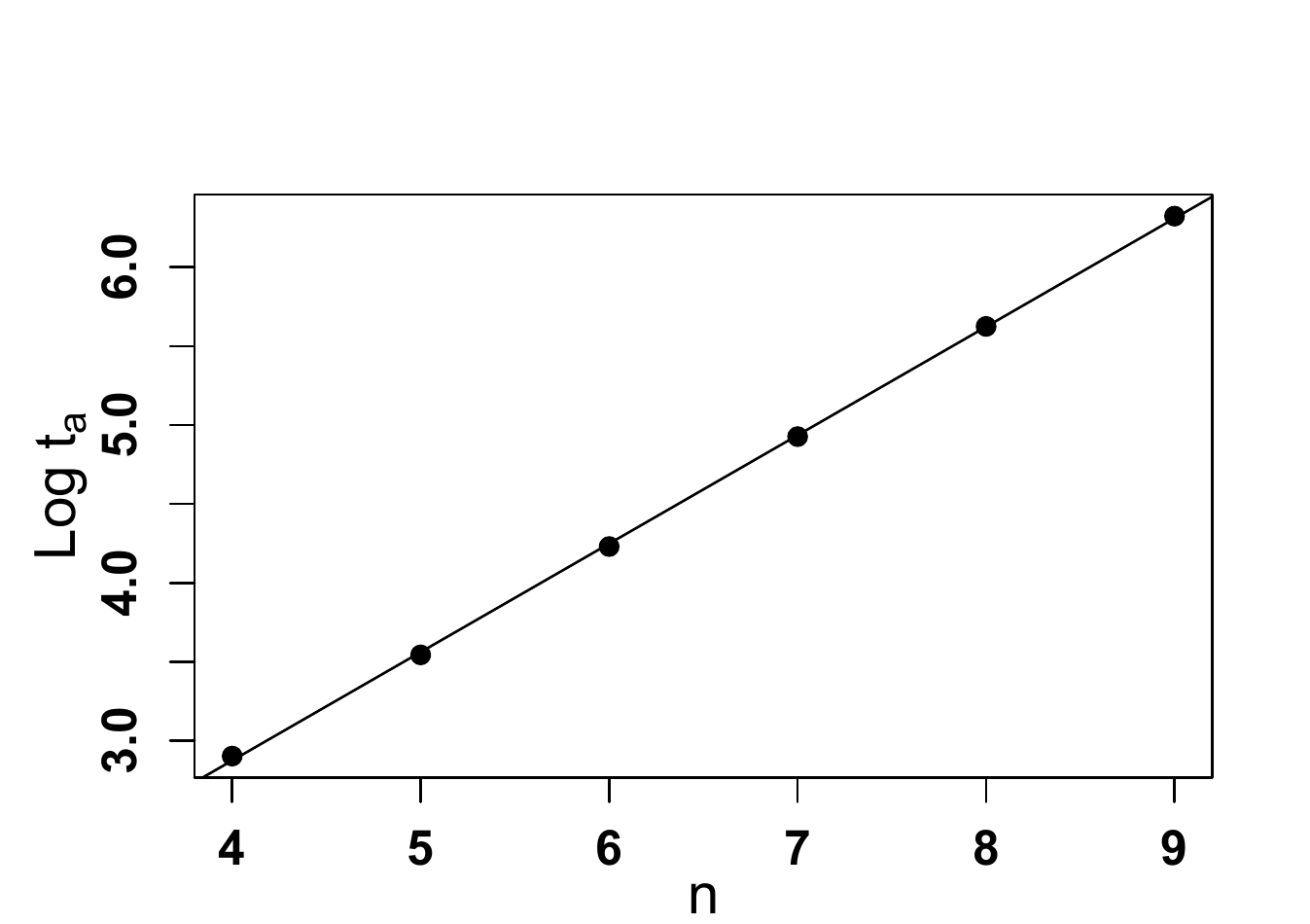}}}}}
\caption{The quasisoliton structure of $ds_n/dt$ as a function of the vehicle index. The fewer the vehicles involved in the quasisoliton, the smaller the width of the quasisoliton, which depends only on $\kappa_0$ and not on the initial headway $s_0$. By convention a kink gives a positively charged quasisoliton as shown in this figure. An anti-kink gives a negatively charged quasisoliton. The top inset shows the dependence of the annihilation time $t_a$ on the number of vehicles between the quasisolitons of opposite charges, the exponential fit is numerically perfect. The intrinsic scale as a function of $s_{c2}$ is shown in the bottom inset.} 
\label{interaction}
\end{figure}

\begin{figure}
  \centering
  \setbox1=\hbox{\includegraphics[height=5.37cm]{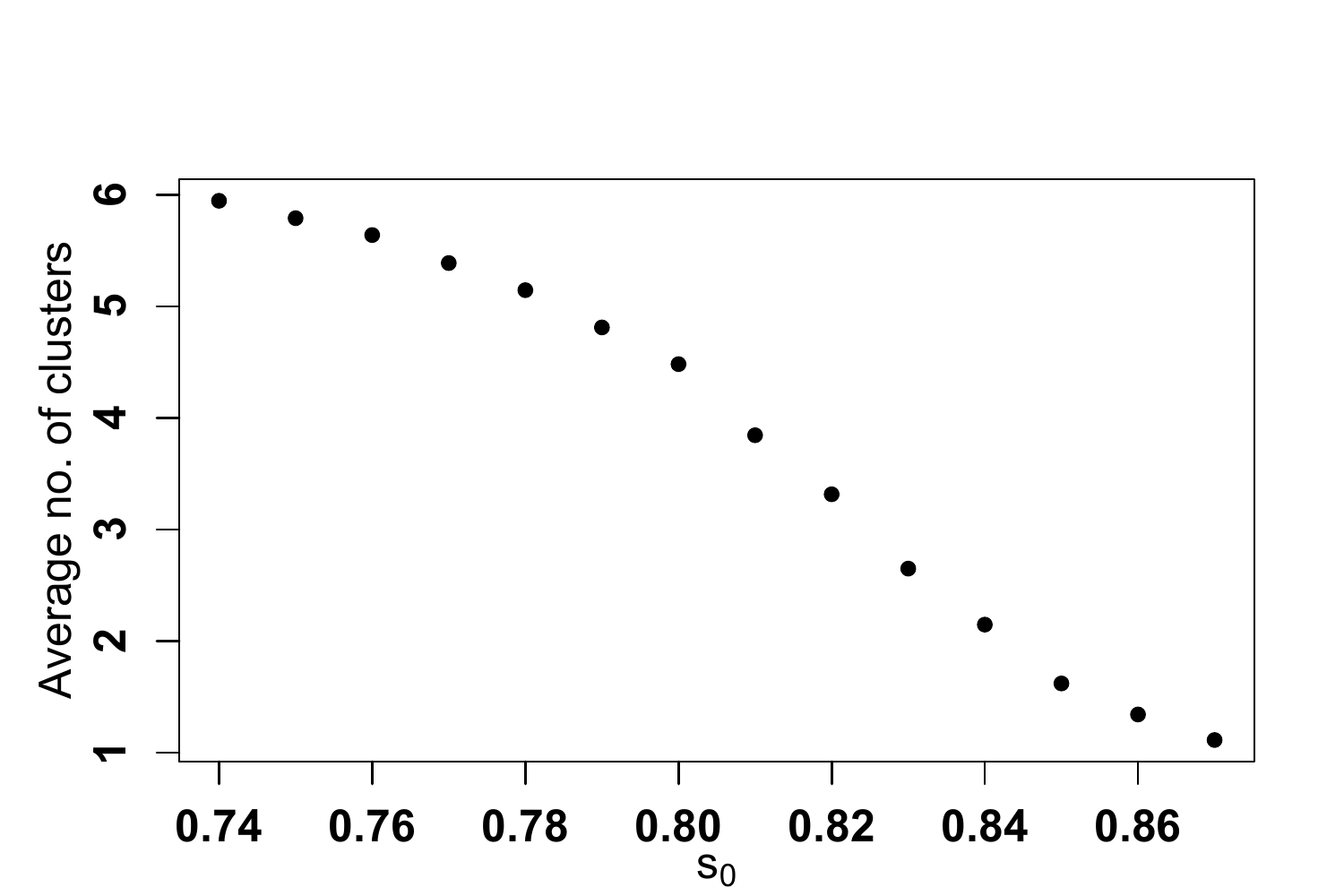}}
\includegraphics[width=8cm]{average1.pdf}{\llap{\makebox[\wd1][l]{\raisebox{-4.5cm}{\includegraphics[height=5.38cm]{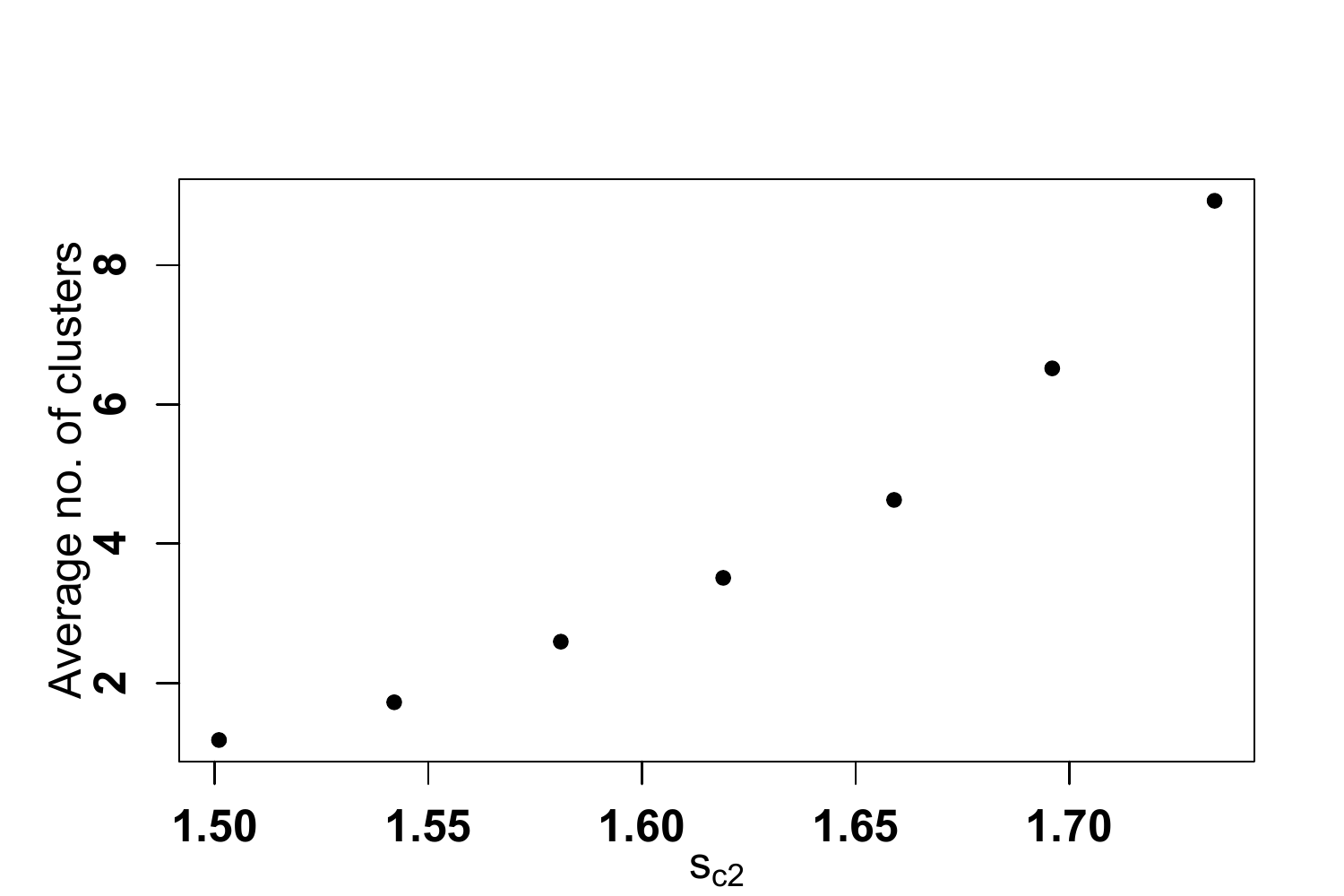}}}}}
\caption{The average number of clusters of a single lane traffic as a function of the intial headway (the top plot, while keeping the perturbation strength and $\kappa_0$ fixed), and as a function of $s_{c2}$ (the bottom plot, while keeping the perturbation strength and $s_0$ fixed).} 
\label{average}
\end{figure}

To understand the probability distribution of the number of clusters, we characterize quantitatively the strength of interaction between two clusters by the time it takes for them to merge. It is useful to plot $ds_n/dt$ instead of $s_n$ as a function of the vehicle index $n$. The ``kinks" and ``anti-kinks" lead to exponentially localized ``quasisolitons" of opposite charges (see Fig.(\ref{interaction})), which closely resemble the ``autosolitons" in dissipative non-linear systems\cite{book}. When quasisolitons of opposite charges annihilate each other, two clusters or anti-clusters merge into one. We numerically observe that the time needed for annihilation, $t_a$,  increases exponentially with the number vehicles $n$ between the peaks of these two quasisolitons, giving the relationship
\begin{eqnarray}\label{ta}
t_a\sim e^{n/n_0}
\end{eqnarray}
One thus note that when $|s_0|$ increases, the cluster (for $s_0>0$) or the anti-cluster (for $s_0<0$) region gets narrower(see Eq.(\ref{jnumber})), leading to higher probability of short distances between quasisolitons. Thus the probability of having multiple (anti-) clusters is suppressed, as shown in Fig.(\ref{probability}). The intrinsic ``scale" $n_0$ in Eq.(\ref{ta}) depends on $s_{c2}$ or $\kappa_0$, which is also plotted in Fig.(\ref{interaction}). This is analogous to the interaction and collapsing of kinks and anti-kinks in the Ginzburg-Landau theory\cite{rougemont}, though here the total number of vehicles in the cluster has to satisfy Eq.(\ref{jnumber}), so that at least one cluster will remain for a finite system with periodic boundary condition. Thus the greater the intrinsic scale, the stronger the interactions between the quasisolitons, so this scale can be used to quantify the absolute value of the quasisoliton charge. The interaction leads to merging of clusters, reducing the probability of having multiple clusters in the traffic lane. Fig.(\ref{probability}) will look qualitatively the same if the x-axis is replaced with increasing $s_{c2}$. The dependence of average number of clusters as a function of $s_0$ and $s_{c2}$ are plotted separately in Fig.(\ref{average}), numerically supporting the above explanation\cite{comment1}.

One should note that the symmetry $s_{max}=-s_{min}=s_{c2}$ is a result from the fact that the RHS of Eq.(\ref{ovmm}) is odd in $s_n$. This symmetry forbids us to tune the corresponding $h_{max}$ and $h_{min}$ independently. The symmetry could be broken when $V_{op}^{(k)}$, or additional terms from $g\left(h_n,\Delta v_n\right)$ in Eq.(\ref{meq2}) contains components that are even in $s_n$. In these cases, all three emergent quantities $s_{max}, s_{min}$ (or the physical headways $h_{max}, h_{min}$) and $n_0$ can be tuned freely with the parameters in the model, providing the necessary degrees of freedom in Sec.~\ref{sec_algorithm} to capture the empirical features in Sec.~\ref{sec_empirical}.

Understanding the multicluster solutions and the physical significance of the intrinsic scale $n_0$ is important in explaining some of the essential features of the traffic dynamics. While spatially random perturbations grow and interact with each other in the metastable and linearly unstable region and eventually form very wide clusters corresponding to the large moving jams, physically $n_0$ characterizes the time scale over which the intermediate transient states can last, as well as the width of the narrow jams that can be detected by the actual measurement. Numerical calculation also shows $n_0$ can be used to tune the maximum acceleration that would normally occur in the congested traffic (which is much smaller than the physical limit of the acceleration of the actual vehicle), making it an important parameter in tuning the effective model.

\section{Algorithm for tuning the traffic model}\label{sec_algorithm}

We will now proceed to construct the simplest effective model, or the minimal model for the real traffic dynamics, that can capture what we observe in Sec.~\ref{sec_empirical}. The discussions in Sec.~\ref{sec_property} is universal, and for a general model given by Eq.(\ref{meqn}), there are also three emergent quantities $h_{max}, h_{min}$ and $n_0$. Physically, $n_0$ also controls the maximum acceleration in the congested traffic, and the interaction between clusters is crucial for the time scale of the evolution of the wide moving jams. We make the following assumptions to start with the simplest possible model from Eq.(\ref{expansion}):
\begin{enumerate}[label=\alph*)]
\item Only the linear order of $\left(\tilde V_{op}^{(k)}-\tilde v_n\right)$ is kept. The model is thus reduced to Eq.(\ref{meq2}).
\item In Eq.(\ref{meq2}) we make both $\kappa_1$ and $g$ independent of the headway $h_n$, so the only headway dependence is within $V_{op}^{(k)}$.
\item We assume $f_0$ decreases monotonically with respect to increasing $v_n$ for any $h_n$ at $\Delta v_n=0$, thus there is only one $V_{op}^{(k)}$.
\end{enumerate}
Following assumption $c)$ we remove the subscript $(k)$ in $V_{op}$. Assumptions in $a)$ and $b)$ are simplifications of the exact model, which can only be justified if $v_n-V_{op}$ is small compared to $V_{max}$, and both $\kappa_1$ and $g$ depend weakly on $h_n$ throughout the time evolution of the traffic dynamics. If we only keep the linear order in $\Delta \tilde v_n$ in our expansion and the expansion coefficient independent of $h_n$, the resulting model is the full force velocity model\cite{JiangR_PRE01}. It does not however contain enough degrees of freedom to capture all the empirical features in Sec.~\ref{sec_empirical}, because of the symmetry $s_{\text{max}}=-s_{\text{min}}$ (see Sec.~\ref{sec_property}). One can either make the coefficient of expansion dependent on $h_n$, or keep the higher orders in $\tilde v_n$. For simplicity we choose the latter option. Here we postulate the empirical features in Sec.~\ref{sec_empirical} can be universally captured by the three emergent quantities of the general traffic models; the simplifications we undertake only remove non-essential microscopic details we are not prepared to capture.

We use the well-studied A-5 North German Highway from Kerner\cite{kernerexp1} as an example. Tuning the model only requires the information from the flow-density plot, which consists of the free flow part (where the flow depends approximately linearly on the density), the congested part (with a collection of randomly scattering data at higher density with suppresed flow) and the wide moving jam given by the ``J line". The list of statistically robust quantities from the flow-density plot we use are:
\begin{eqnarray}\label{exp}
&&V_{\max}=\lim_{\rho\rightarrow 0}dQ/d\rho\sim 42ms^{-1}, \quad Q_{\text{max}}\sim 3000 veh/h,\nonumber\\ &&\quad\rho_{\text{c}}\sim 30 veh/km, \quad Q_{dj}\sim 2000 veh/h, \nonumber\\
&&\quad\rho_{dj}\sim 17.5 veh/km, \quad\rho_j\sim 125 veh/km
\end{eqnarray}
Here $\rho_{\text{c}}$ is the critical density of the highway at which $Q_{\text{max}}$, the maximum flow, is observed. $Q_{dj}$ and $\rho_{dj}$ are the flow and density downstream of the wide moving jam respectively, while $\rho_j$ is the density within the jam. For the lack of the raw traffic data, all the numerical values are rough estimates only, and for our purpose of illustration that is sufficient, as we do not need to fine-tune the model to simulate the qualitative empirical features. We also assume on average the length of the vehicle $l_c=5m$, and by identifying the parameters of the cluster structure with the characteristic parameters of a wide moving jam we have the following relationship:
\begin{eqnarray}\label{relationship}
&&V_{op}\left(\infty\right)=V_{max}, \quad V_{op}\left(h_{max}\right)=v_{dj}=Q_{dj}/\rho_{dj}\nonumber\\
&&V_{op}\left(h_{min}\right)=0, \quad V_{op}\left(h_{cr}\right)=v_{cr}=Q_{\text{max}}/\rho_{c}
\end{eqnarray}
The two other characteristic velocities from the flow-density plot are $V_j=Q_{\text{max}}/\left(\rho_{c}-\rho_j\right)$, the velocity of the downstream front of a wide moving jam, and $V_C=\left(Q_{\text{max}}-Q_{dj}\right)/\left(\rho_{cr}-\rho_{dj}\right)$, the velocity of the downstream front between $Q_{\text{max}}$ and $Q_{dj}$. The cluster parameters are given by $h_{max}=\rho_{dj}^{-1}-l_c, h_{min}=\rho_j^{-1}-l_c, h_{cr}=\rho_{c}^{-1}-l_c$.

The simplest form of $V_{op}$ has been suggested in\cite{helbing6} for its analytic tractability; it however does not capture the correct fundamental diagram in the free flow phase. Here we solve Eq.(\ref{relationship}) most simply with a piecewise function passing through $\left(h_{min}, 0\right), \left(h_{cr}, v_{cr}\right), \left(h_{max}, v_{dj}\right)$ and bounded at $V_{max}$, so as to fix the quantitative features of the real traffic dynamics\cite{footnote2} in the free flow phase. Defining $h_c=\frac{V_{max}-V_C}{v_{cr}-v_{dj}}\left(h_{cr}-h_{max}\right)-l_c$ we have:
\small
\begin{eqnarray}\label{piecewise}
V_{op}\left(h\right)=\left\{
\begin{array}{lr}
0 &  h<h_{min}\\
\frac{v_{cr}}{h_{cr}-h_{min}}\left(h+l_c\right)+V_j & h_{cr}>h\ge h_{min}\\
\frac{v_{cr}-v_{dj}}{h_{cr}-h_{max}}\left(h+l_c\right)+V_C& h_c>h\ge h_{cr}\\
V_{max} & h\ge h_c
\end{array}
\right.
\end{eqnarray}
\normalsize
 
Thus the fundamental diagram is defined by $V_{op}$ (see Fig.(\ref{fig2b})), capturing the quantitative features of the empirical flow-density plot. The next step is to tune $\kappa_0$ and $g\left(\Delta v_n\right)$ so that the cluster solutions have the desirable $h_{max}, h_{min}$ and $n_0$. We adopt the reasonable assumption that the maximum acceleration for the vehicles in the stop-and-go wave should be within the range of $\pm 3ms^{-2}$. Small variations around this quantitative assumptions do not qualitative alter the arguments and conclusions in this paper. Given that three parameters need to be fixed, $g\left(\Delta v_n\right)$ has to be non-linear and contain terms that are even in $\Delta v_n$ to break the symmetry of $s_{max}=-s_{min}$ (see Sec.~\ref{sec_property}). We choose to simply adopt the AFVD model\cite{GLW_PhyA08} (in the case where $\lambda_2\ne 0$) as follows
\begin{eqnarray}\label{g}
g\left(\Delta v_n\right)=\lambda_1\Delta v_n+\lambda_2|\Delta v_n|
\end{eqnarray}
The effective model is now completely defined, and with numerical calculations the fitted parameters are $\kappa_0=0.1s^{-1}, \lambda_1=6.2$ and $\lambda_2=-2.9$, corresponding to $h_{min}\sim 3m$ and $h_{max}\sim 52m$. 

A summary of our algorithm is in order. The optimal velocity function we chose defines the fundamental diagram: it gives the correct maximum average velocity of the traffic system (when the density  of the traffic on the highway is very small). It also gives the right $(\rho_{dj}, Q_{dj})$ and $(\rho_c, Q_{\text{max}})$ pairs on the flow-density plot, where the traffic is still in the free flow phase. In addition, it gives the maximum density $\rho_j$ of the traffic. While these are just some of the special points on the flow density plot, the tuning of the other parameters in the model ($\kappa_0, \lambda_1,\lambda_2)$ makes sure $(Q_{dj},\rho_{dj})$ corresponds to the characteristic flow and density downstream of a wide moving jam, and $\rho_j$ corresponds to the density within a wide moving jam. In addition, it also makes sure the acceleration of vehicles in congested traffic is physically reasonable.

The parameters in the model all have very clear physical meanings; the general features of the optimal velocity function is also quite intuitive. Both $V_{op}\left(h_n\right)$ and $g\left(\Delta v_n\right)$ are not smooth, which is not physical. They are, however, just unimportant artifects of the model that can be easily (but tediously) removed mathematically without affecting any of the conclusions or the predictive powers of the model. One should note that the optimal velocity function and the parameters in the model do not explicity tell us about the transition from the free flow to the congested flow, or the maximum flow that can be achieved, as well as the qualitative features of the complex spatiotemporal patterns. All these features will be predicted by the dynamics of the model which we will show in the following section.

\section{Predictions of the effective model}\label{sec_prediction}

\begin{figure}
  \centering
  \setbox1=\hbox{\includegraphics[height=4.6cm]{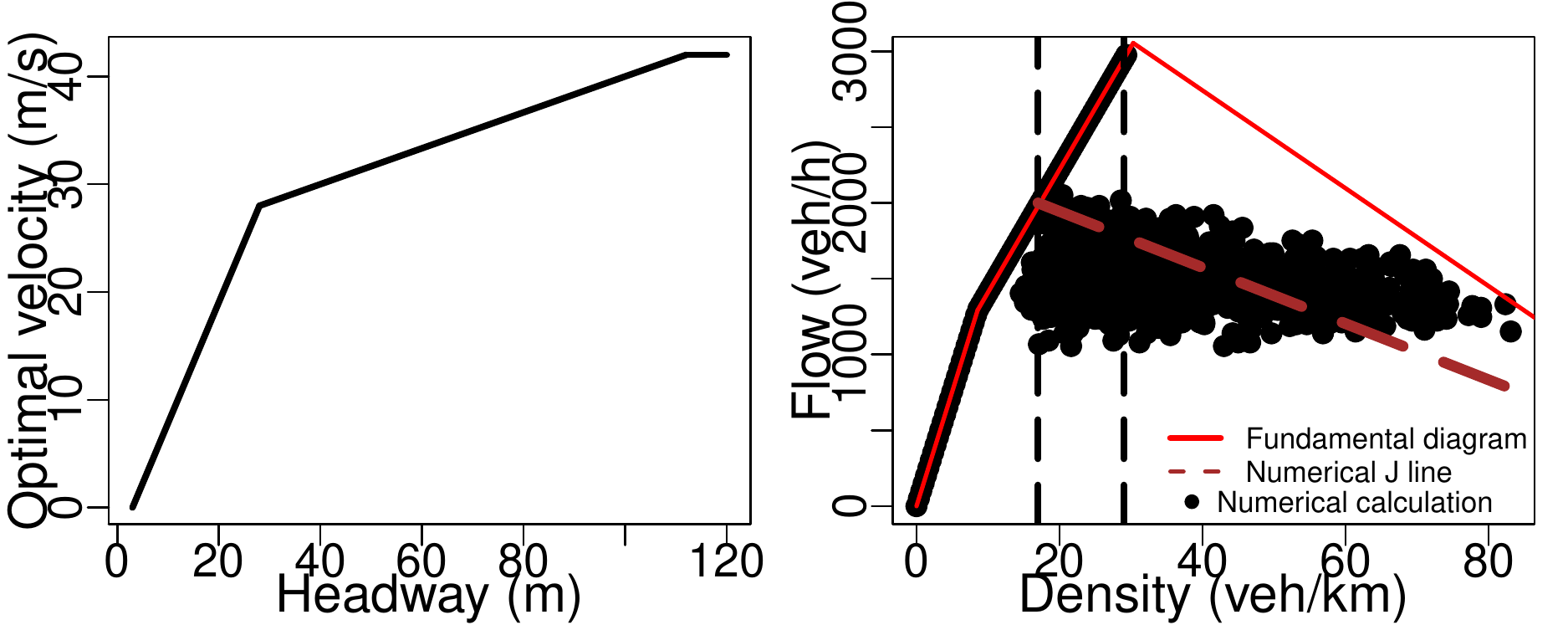}}
  \includegraphics[height=3.5cm]{fig2b.pdf}{\llap{\makebox[\wd1][c]{\raisebox{0.65cm}{\includegraphics[height=1.8cm]{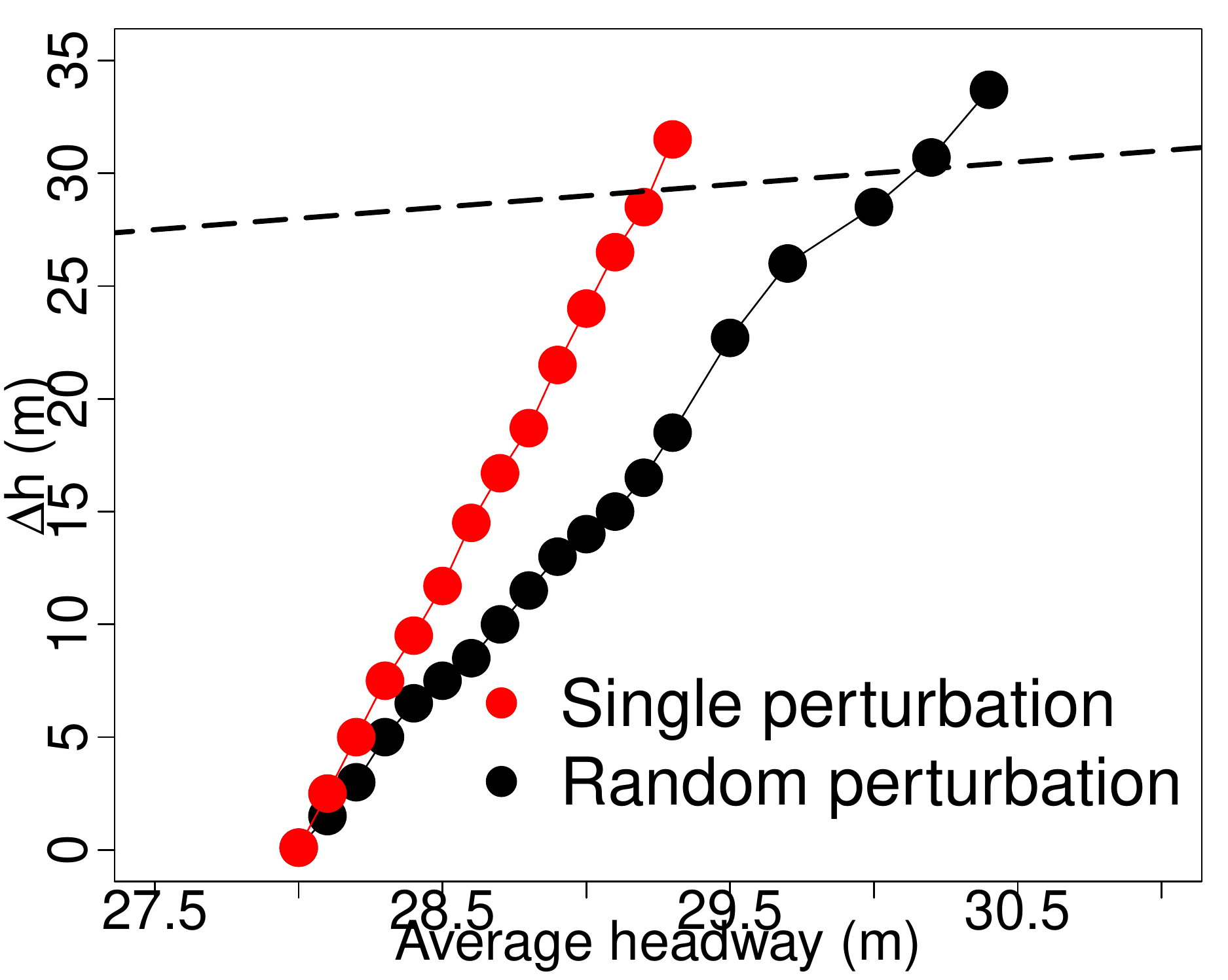}}}}}
  \caption{(Color online) Characteristics of the effective OV model. Left: The piecewise optimal velocity function. Left inset: The magnitude of perturbation ($\Delta h$) needed to form cluster solutions in the metastable region. Cases for the single vehicle perturbation (red) and the random perturbation (black) are plotted. Right: The actual flow-density diagram including both the free flow and the congested flow. The plot is obtained from the numerical calcuation with open boundary condition and an on-ramp bottleneck. The left vertical line gives $\rho_{dj}$ and the right vertical line gives $\rho_{cr}$.}
\label{fig2b}
\end{figure}

We now proceed to examine what the minimal model predicts about the traffic dynamics. Given the parameters in the model, the free flow in the stable phase is given in the region $\rho<\left(h_{max}+l\right)^{-1}\sim 17veh/h$. The metastable region is given by $17veh/km\lesssim\rho\lesssim 30veh/km$. In this region, a large enough perturbation will grow in time and leads to the instability of the free flow and formation of the jams. The empirical feature that the free flow persists up to the critical density $\rho_{c}\sim 30 veh/h$ is nicely predicted by the fact that for the metastable region with density smaller than $\sim 30 veh/h$, the perturbation needs to be greater than the average vehicle headway for the free flow to be unstable (see Fig.(\ref{fig2b})), which is unlikely without collisions. Thus the effective model captures $Q_{\text{max}}$ and $\rho_{c}$ quite accurately, even though Eq.(\ref{piecewise}) in no way guarantee the stability condition agreeing with the empirical data.

\begin{figure}
\includegraphics[width=8.5cm]{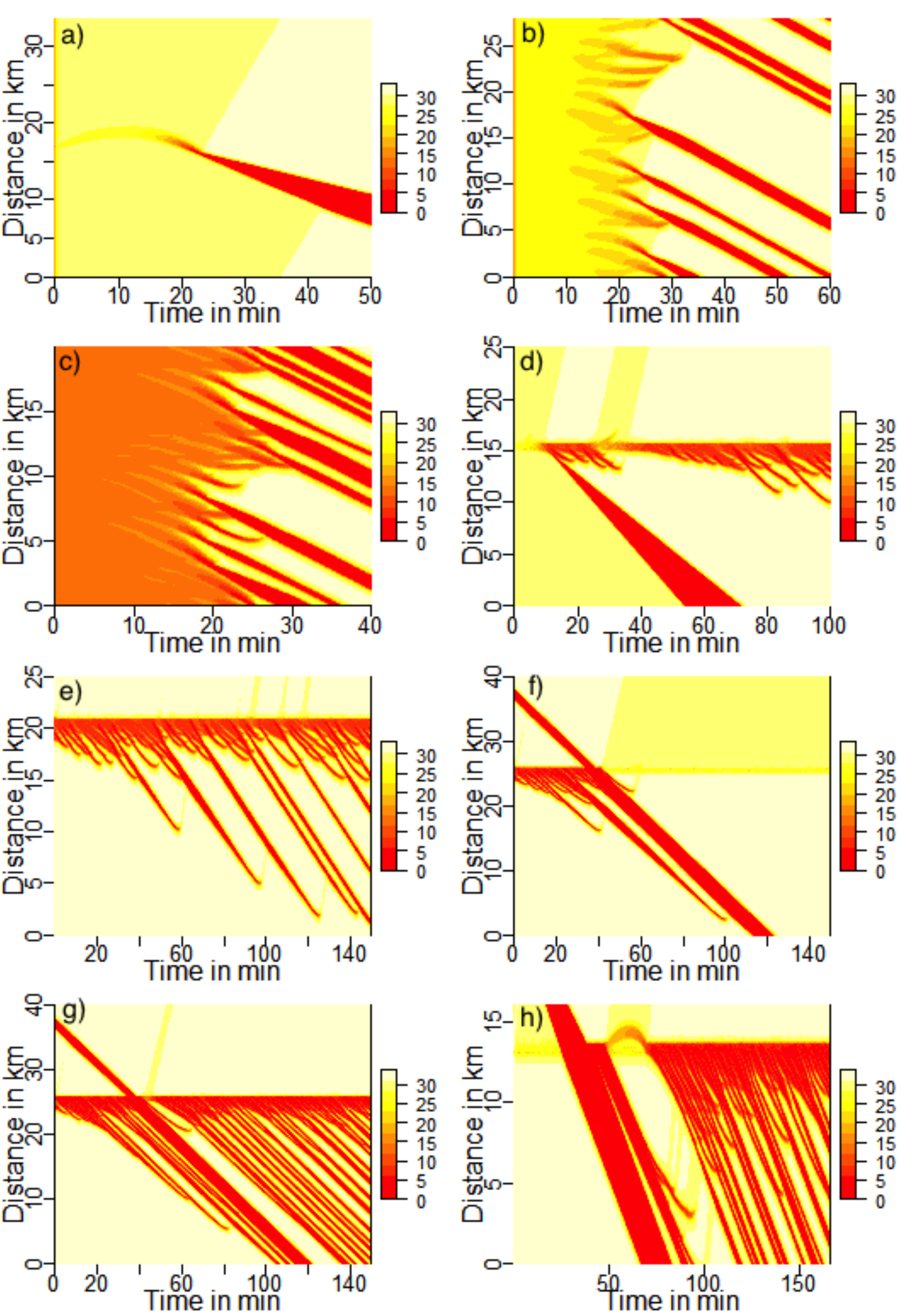}
\caption{(Color online) Various traffic patterns from the minimal effective model with the velocity color plot. Plot a)$\sim$c) are numerical calculations from the periodic boundary condition; the rest are from the open boundary condition with the presence of an on-ramp bottleneck. a). The ``boomerang" behavior with average density close to the coexistence curve, which starts at t =0 min at around distance $\sim$ 16 km. b). The ``pinch effect" in the metastable region. c). Numerous narrow moving jams in the linearly unstable region.  d). A wide moving jam followed by the congested traffic at the bottleneck when the main traffic density is metastable. e). The formation of the congested traffic at the bottleneck when the main traffic density is stable. f). Small $Q_{in}$ with metastable main traffic flow, a wide moving jam effectively stops the congested traffic because the density downstream of the jam is lower than the that of the main traffic. g). A wide moving jam passes through the bottleneck with its downstream front maintaining its characteristic velocity. h). A wide moving jam induces congested traffic at the bottleneck. The main traffic density is lower than the density downstream of the wide moving jam.} 
\label{fig3b}
\end{figure}

To study the congested traffic and the evolution of the wide moving jams, both periodic boundary condition of a single homogeneous lane and open boundary condition of a single lane with the presence of an on-ramp bottleneck are simulated. For periodic boundary conditions, the initial condition is chosen to have random fluctuations of the headways with different average density. Though in the long time limit wide moving jams eventually form for average density $\rho\gtrsim 29 veh/km$, the intermediate process can be quite complicated. When the average density is very close to the phase boundary, or the coexistence curve given by $\left(h_{max}+l\right)^{-1}$, very large perturbations are needed to nucleate a wide moving jam via the well-known ``boomerang" behavior\cite{helbing1,kernerbook}; when the average density increases further, the ``pinch effect"\cite{kernerbook} is observed at multiple locations leading to multiple narrow jams; at high density numerous narrow jams form relatively quickly, and over time these narrow jams interact and merge into a few wide moving jams (see Fig.(\ref{fig3b})).

While the ``pinch effect" and the formation of numerous narrow jams are well-known in the literature\cite{kernerbook}, the existence of the ``boomerang" behavior is still debated\cite{kernerbook,helbing1}. The absence of ``boomerang" behavior may also due to the rarity of very large perturbations on a multi-lane highway, when accidents and bottlenecks are absent. Even for moderately large perturbations, it takes more than an hour for a closed traffic system to develop the ``boomerang behavior". One should also note that based on the numerical simulation, it takes $30\sim 60$ minutes for the wide moving jams to eventually emerge from a random initial condition via complex intermediate states. Thus comparing numerical results with a fixed number of vehicles and periodic boundary condition to the empirical observations can be extremely tricky. The real world traffic, being an open system, does not maintain its vehicle density and the total number of vehicles over an extended period of time; variation of the average density within the metastable/unstable region leads to a mixture of long lasting intermediate states, numerous narrow jams and occasional wide moving jams. 

From both the empirical validation and transportation engineering points of view, numerical simulations of the effective model in the presence of bottlenecks are more crucial. In our simulation with open boundary condition the virtual sensor measures the flow and average velocity of the passing vehicles in \emph{exactly} the same way as the traffic sensors installed in the real world highways\cite{helbing2}. We use the idealized initial condition with a constant main traffic flow $Q_m$ and an on-ramp flow $Q_{in}$. At low enough $Q_{in}$ the free flow is maintained, though in the linearly unstable region the on-ramp flow has to be close to zero. When $Q_{in}$ increases, corresponding to the increase in the bottleneck strength, the congested flow develops immediately upstream of the bottleneck. This region with length $L_c$  share the characteristics of the ``synchronized flow" (see Fig.(\ref{fig2b}), Fig.(\ref{fig3b})). Narrow jams form upstream of the congested flow, and wide moving jams appear upstream of these narrow jams from merging of the narrow jams and the ``pinch effect". The ``boomerang" behavior is also observed (see Fig.(\ref{fig3b}h)). The congested flow can be either spontaneous or induced by a passing wide moving jam. In general, $L_c$ can be as long as $4km$ and decreases with the increase of $Q_{in}$. 

\begin{figure}
\includegraphics[width=8cm]{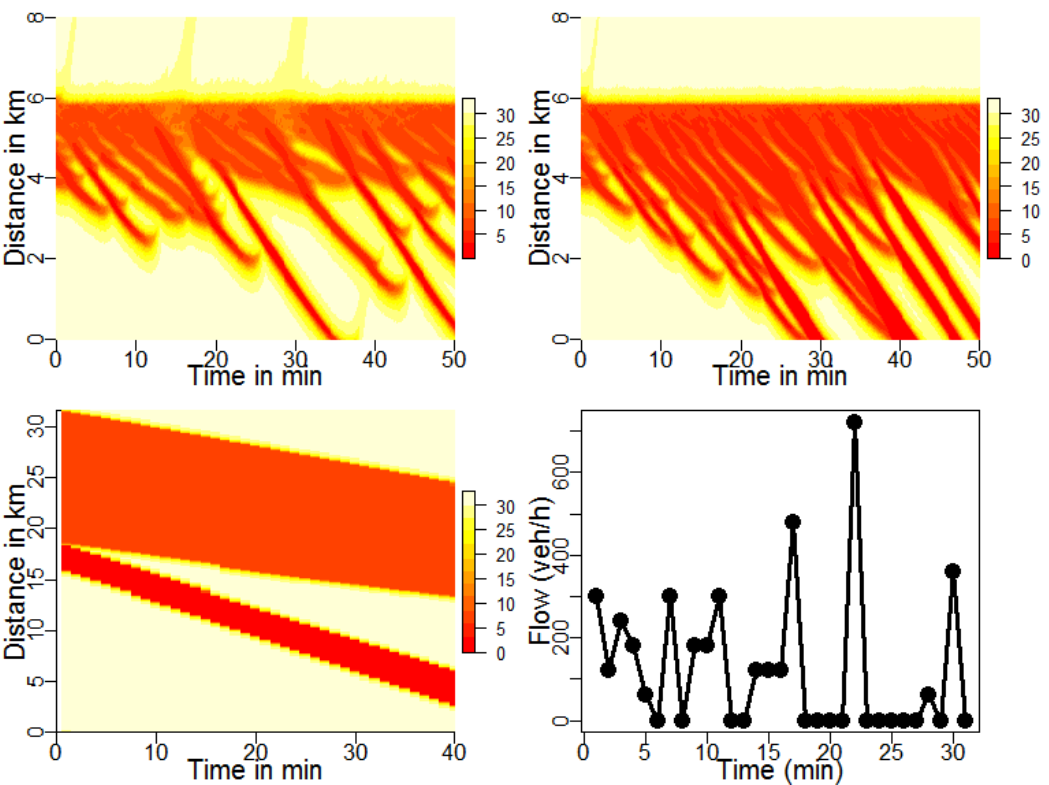}
\caption{(Color online) Additional features predicted by the minimal effective model in Sec.~\ref{sec_algorithm} with the velocity color plot except the bottom right one. Top left: Congested pattern at the bottleneck with $Q_m=1800 veh/h$ and $Q_{in}=360 veh/h$. Top right: Congested pattern at the bottleneck with $Q_m=1800 veh/h$ and $Q_{in}=500 veh/h$. Both the narrow jams and the wide moving jams emerge closer to the bottleneck, and fluctuating with higher frequencies. Bottom left: Dynamics of a model state with the wide moving jam adjacent to the homogeneous congested state. The velocity of the downstream front of the wide moving jam is unaffected by the traffic conditions downstream of the jam. Bottom right: The flow time series when the traffic density is $80 veh/km$ } 
\label{fig4b}
\end{figure}

Previous GM based models were criticized in \cite{kernerbook} based on some fundamental empirical observations of the congested phase at the bottleneck, as well as on the absence of the homogeneous congested traffic (HCT) empirically. In contrast, the effective model we constructed agrees with the empirical observation that increasing the bottleneck strength leads to higher frequency of moving jam emergence and smaller $L_c$ (see Fig.(\ref{fig4b})). In fact this is the most common situation for various different $Q_m$. Numerical calculations also show the metastable phase in the high density region is very narrow. The model actually predicts complicated spatio-temporal structures for the congested traffic at very high density, with traffic flow fluctuate between zero to 500 $veh/h$, agreeing with the empirical observation in\cite{kernerbook}. This is simply because small perturbation is linearly unstable even in the region of vehicle density up to $\sim 100 veh/km$.

\section{Summary and Outlook}\label{sec_conclusion}

In summary, we have presented a systematic way of constructing traffic models that in principle can capture all the empirical features for which stochasticity as well as diversity in drivers and vehicle types \emph{do not} play a fundamental role; only the averaged non-linear interaction between the vehicles is fundamental. The exact form of the traffic model given by Eq.(\ref{meqn}) can be obtained by the empirical measurement. In the case that the model is analytic, one should obtain various effective models by expansions around the ground states defined in Sec.~\ref{sec_model}. The compromise between keeping the model simple and capturing more microscopic details can now be done in a systematic way by choosing the appropriate set of parameters to average over, and by making various approximations in the expansions. 

In addition, we proposed a simple algorithm to justify the approximations we made by expanding the traffic model around a unique ground state.  The resulting minimal effective model shows that the physics of many empirical observations of the highway traffic dynamics can be captured by a deterministic effective model based on a simple optimal velocity function we proposed. In this framework the congested traffic is characterized by long lasting transient states of the model, from which the wide moving jams evolve from the ``pinch effect" or the merging of the narrow jams. Interestingly, our results imply it is probably difficult to distinguish between real traffic dynamics and those from identical autonomous vehicles with simple driving rules based on the empirical data, including the flow-density plot and the congestion patterns near the bottlenecks. One should note that even deterministic models can simulate seemingly unpredictable dynamics because in reality the initial condition can be random and the highway traffic is an open system. 

While the minimal effective model we constructed give convincing evidence that many complex traffic dynamics can still be explained \emph{solely} by the simple interactions between nearest neighbour vehicles, we would not expect the model to explain all the interesting empirical features\cite{helbing}. With more advanced techniques in collecting and analyzing empirical data, more empirical features can be properly defined, and we would expect to relax certain assumptions in our effective model(see Table.~\ref{tt}) to capture those features. One should note that the optimal velocity model class emerge naturally from an analytic traffic model, though such model can still be complicated if there is more than one ground state for a fixed average density. This, together with the possibility that the model could be non-analytic within certain density range, can justify the parameter rich ``three-phase models" from a more fundamental ground. A long term, microscopic measurement of the vehicle acceleration as a result of the interaction between its close neighbours under diverse environmental settings is currently work in progress. The empirical data, together with proper averaging, should be able to give us Eq.(\ref{meqn}) and thus provide insight on which effective model is more consistent with the nature.
\begin{acknowledgements}
We would like to thank Prof. Weizhu Bao and Prof. Ren Weiqing from National University of Singapore for useful comments. This research was partially supported by Singapore A$^{\star}$STAR SERC ``Complex Systems" Research Programme grant 1224504056. The numerical calculations in this work is supported by ACRC of A$^{\star}$STAR. 
\end{acknowledgements}


\begin{thebibliography}{99}
\bibitem{footnote1}
Interactions with vehicles further ahead or at the back are generally small and will not change qualitatively the conclusions in this work, see for example \cite{xue}. We can also include the effects of vehicles further away into $\{n\}$ in Eq.(\ref{acc}), without loss of generality.
\bibitem{footnote2}
The most popular form of $V_{op}$ used in the literature is $V_1+V_2\tanh\left(C_1h-C_2\right)$, due to its mathematical simplicity, but not necessarily due to its closeness to the real $\bar f$. While the form gives enough degrees of freedom, it does not have a solution based on the empirical data in Eq.(\ref{exp}). Eq.(\ref{piecewise}) is a simple and straightward way of constructing $V_{op}$ based on any reasonable empirical data.
\bibitem{comment1}
For any finite number of vehicles, all clusters will eventually merge in the limit of very long time; thus the statements here are only rigorous in the limit that the number of vehicles $N\rightarrow 0$. However because of the exponential dependence of the annihilation time on the number of vehicles between quasisolitons of opposite charges, the statements here are true for all practical purposes when the number of vehicles is reasonably large (even for computer simulation because of the finite numerical resolutions)
\end{thebibliography}
\end{document}